\documentclass[12pt]{article} 
\usepackage{graphicx}
\usepackage{epsfig}

\def\(({\left(}
 \def\)){\right)}
\def\[[{\left[}
\def\]]{\right]}
\newcommand {\artanh} {\mbox{artanh}}

\newcommand {\bi} {\bibitem}
\newcommand {\be} {\begin{equation}}
\newcommand {\ee} {\end{equation}}
\newcommand {\bea} {\begin{eqnarray} }
\newcommand {\eea} {\nonumber \end{eqnarray} }
\newcommand {\eps} {\epsilon}
 \newcommand {\si} {\sigma}

 \newcommand {\al} {\alpha}

\newcommand {\ba} {\overline}
\newcommand {\lan} {\langle}
\newcommand {\ran} {\rangle}

\newcommand {\cD}  {{\cal D}}

\newcommand {\cN}  {{\cal N}}

\newcommand {\cP}  {{\cal P}}

\newcommand {\bc} {\begin{center}}
\newcommand {\ec} {\end{center}}
\newcommand {\bd}{\begin{displaymath}}
\newcommand {\ed}{\end{displaymath}}

\newcommand {\tgh} {\mbox{th}}
\newcommand {\arth} {\mbox{arth}}

\def \form#1 {eq. (\ref{#1}) }
\def \parziale#1#2  {{\partial {#1} \over \partial {#2}}}
\def \bi#1 {\typeout{#1} \item}

\begin{document}
\title{Mean field theory of spin glasses: statics and dynamics}

\author{Giorgio Parisi\\
Dipartimento di Fisica, Sezione INFN, SMC-INFM,\\
Universit\`a di Roma ``La Sapienza'',\\
Piazzale Aldo Moro 2,
I-00185 Rome (Italy)\\
Giorgio.Parisi@Roma1.Infn.It}
\maketitle

\begin{abstract}
	 In these lectures I will review some theoretical results that have been obtained  for
	 spin glasses.  I will concentrate my attention on the 
	 formulation of  the
	 mean field approach and on its numerical and experimental verifications.  I will present  the
	 various hypothesis at the basis of the theory and I will discuss their mathematical and physical
	 status.
\end{abstract}

\section{Introduction}
Spin glasses have been intensively studied in the last thirty years. They are very interesting for many
reasons:
\begin{itemize}
\item
Spin glasses are the simplest example of glassy systems.  There is a highly non-trivial mean field
approximation where one can study phenomena that have been discovered for the first time in this
context, e.g. the existence of many equilibrium states and  the ultrametricity relation among these states.
In this framework it is possible to derive some of the main properties of a generic glassy systems, e.g.
history dependent response \cite{ANDERSON,MPV,PARISIB}; this property, in the context of mean field
approximation, is related to the existence of many {\sl equilibrium states}~\footnote{This sentence is too
vague: one should discuss the its precise  mathematical meaning; although we
will present in these lecture a physically reasonable definition, for a careful discussion see ref.
\cite{NS,I,CINQUE,PAR1,PAR2}.}.
\item
The study of spin glasses opens a very important window for studying 
off-equilibrium behavior.  Aging \cite{Bou} and the related violations of the equilibrium fluctuation
dissipation relations emerge in a natural way and they can be studied in this simple setting
\cite{Cuku1,Cuku2,FM,FMPP}.  Many of the ideas developed in this context can be used in other physical fields
like fragile glasses, colloids, granular materials, combinatorial optimization problems and for other
complex systems \cite{MM}.
\item
The theoretical concepts and the tools developed in the study of spin glasses are based on two
logically equivalent,
but very different, methods: the algebraic broken replica symmetry method and the probabilistic cavity
approach \cite{MPV,PARISIB}. They  have a wide domain of applications. Some of the properties that appear in the
mean field approximation, like ultrametricity, are unexpected and  counterintuitive. \item
Spin glasses also provide a testing ground for a more mathematical inclined probabilistic approach: the
rigorous proof of the correctness of the solution of the mean field model came out after twenty years of
efforts where new ideas (e.g. stochastic stability \cite{SS1,SS2,SS3}) and new variational principles
\cite{V1,V2} were at the basis of a recent rigorous proof \cite{TALA} of the correctness of the mean field
approximation for in the case of the infinite range Sherrington-Kirkpatrick model \cite{SK}.
\end{itemize}

In these lectures  I will present a short review of some of the results that have been obtained using
the  probabilistic cavity approach. I will mostly discuss the mean field approximation for the infinite range Sherrington-Kirkpatrick.
I will only mention {\em en passant} how to extend these results to finite connectivity model and to finite
dimensional systems. The very interesting applications of these techniques to other problems coming from
physics (e.g. glasses \cite{LH2002}) and other disciplines \cite{MM} will not be discussed.

\section{General considerations}

The simplest spin glass Hamiltonian is of the form:

\be H=\sum_{i,k=1,N}J_{i,k} \si_{i}\si_{k}\, ,  \label{SIMPLEST}
\ee 
where the $J$'s are {\it quenched} (i.e. time independent) random variables located on the links  connecting
two points of the lattice and
the $\si$'s are Ising variables (i.e. they are equal to $\pm 1$).  The total number of points is denoted with $N$ and
it goes to infinity in the thermodynamic limit.

We can consider four models of increasing complexity:
\begin{itemize}
\item
The SK model \cite{SK}: All $J$'s are random and different from zero, with a Gaussian or a bimodal
distribution with variance $N^{-1/2}$.  The coordination number ($z$) is $N-1$ and it goes to infinity with
$N$.  In this case a \emph{simple} mean field theory is valid in the infinite $N$ limit \cite{MPV,PARISIB}.
\item
The Bethe lattice model \cite{VB,MP1Be,MP2Be}: The spins live on a random lattice and only $Nz/2$ $J$'s are
different from zero: they have variance $z^{-1/2}$.  The average coordination number is finite
(i.e. $z$).  In this case a modified mean field theory is valid.
\item
The large range  Edwards Anderson model \cite{FL}: The spins belong to a finite dimensional lattice
of dimension $D$.  Only nearest spins at a distance less than $R$ interact and the variance of the $J$'s is
proportional to $1/R^{D/2}$.  If  $R$ is large,   the corrections to mean field theory are small for
thermodynamic quantities, although they
may change the large distance behavior of the correlations functions and the nature of the phase transition,
which may disappear.

\item
The Edwards Anderson model \cite{EA}: The spins belong to a finite dimensional lattice of dimensions
$D$: Only nearest neighbor interactions are different from zero and their variance is $D^{-1/2}$.
In this case finite corrections to mean field theory are present, that are certainly very large in
one or two dimensions, where no transition is expected. The Edwards Anderson model  corresponds to the limit $R=1$ of the 
the large range  Edwards Anderson model; both models are expected to belong to the same universality class. The
large range model Edwards Anderson provides a systematic way to interpolate between the mean field results and
the short range model~\footnote{In a different approach one introduces $\nu$ spins per site that are coupled
to all the spins in the nearest points of the lattice. It is possible formally to construct an expansion in the parameter
$g\equiv 1/\nu$, the so called loop expansion.}. 

\end{itemize}

As far as the free energy is concerned, one can prove the following  rigorous results:
\bea
\lim_{z \to \infty}\mbox{Bethe}(z)=\mbox{SK} \ , \\
\lim_{D \to \infty}\mbox{Edwards Anderson}(D)=\mbox{SK}\ , \nonumber \\
\lim_{R \to \infty}\mbox{finite range EA}(R)=\mbox{SK}\ .
\eea

The Sherrington Kirkpatrick model is thus a good starting point for studying also the finite
dimensional case with short range interaction, that is the most realistic and difficult case. This starting
point becomes worse and worse when the dimension decreases, e.g.  it is not any more useful in the limiting 
case where $D=1$.
\section{Mean field theory}
\subsection{General considerations}
Let us start again with the Hamiltonian in eq. (\ref{SIMPLEST})
and let us proceed in the most naive way. Further refinements will be 
added later.

We consider the local magnetizations $m_{i}=\lan \sigma_{i}\ran$ and we write the simplest possible
form of the free energy as function of the magnetization.
Using the standard arguments, that can be found in  books \cite{I}, we get:
\be
F[m]=\sum_{i,k=1,N}J_{i,k} m_{i}m_{k} - T \sum_{i}S(m_{i}) \ ,
\ee
where $S(m)$ is the usual binary entropy:
\be
-S(m) =
\frac{1}{2}\left( \frac{1+m}{2} \log\left(\frac{1+m}{2}\right)+
\frac{1-m}{2}\log\left(\frac{1-m}{2}\right)\right) \ .
\ee
If we look to the minimum of the free energy, we get the following equations (that are valid at any
stationary point of the free energy):
\be
m_{i}=\tgh(h^{eff}_{i}),\ \ \ \ \ \ h^{eff}_{i}=\sum_{k}J_{i,k}\sigma_{k} \label {NAIVE} \ .
\ee

This is well known and fifty years ago we could have stopped here. However now we understand that 
mean field approximation involves uncontrolled approximations and therefore we need to work in a more
controlled framework. 

In a more modern approach one
consider a model  and one tries to write down an expression for
the free energy that is exact for that particular model. In this way one is sure that the range of
validity of the formulae one is writing is not void (and what is more important, if no technical mistakes have
been done, there could be no contradictions). As far as the exact model is often obtained as limit
of a more realistic model when some parameter goes to infinity, it is possible to estimate the
corrections to these asymptotic results.

A very interesting case, where usually mean field exact formula are valid, happens when the coordination
number (i.e. the number of spins that are connected to a given spin) goes to infinity.  Let us consider the
following construction.  We assume that for a given $i$, $J_{i,k}$ is different from zero only for $z$
different values of $k$ (for simplicity let us also assume that the $J$'s  take only the values $\pm 1$ with
equal probability).

We are interested to study the limit where $z$ goes to infinity.  One can immediately see that in the random
case a finite result
is obtained (at least in the high temperature phase) only if $J$ goes to zero as $z^{-1/2}$ (a similar result
can be obtained also in the low temperature phase).  This result is in variance with the ferromagnetic case
where the $J$'s are all positive and they should go zero as $z$ in order to have a finite result in the low
temperature phase.  Indeed a good thermodynamic scaling is present when $h^{eff}$ is of O(1).  In the
ferromagnetic case the $z$ terms in the expression for $h^{eff}$ add coherently and therefore each of them
must be of order $z^{-1}$; on the contrary in the spin glass case, if they are uncorrelated (this is true in a
first approximation), $h^{eff}$ is the sum of $z$ random terms and the result is of O(1) only if each term is
of order $z^{-1/2}$.

If one makes a more careful computation and we look to the corrections to the mean field  expression that
 survive in this limit, one obtains:
\bea
F_{TAP}[m]=\\
\sum_{i,k=1,N}\left(\frac12 J_{i,k} m_{i}m_{k}
-\frac14\beta J_{i,k}^{2} (1-m_{i}^{2})(1-m_{k}^{2})\right)- T \sum_{i}S(m_{i})\ ,
\eea
where $\beta=1/T$ (we put the  Boltzmann-Drude constant $\alpha$  to the value 3/2). 

This free energy has an extra term with respect to the previous free energy eq. (\ref{NAIVE}) and is called the TAP free energy
\cite{TAP}, because it firstly appeared in a preprint signed by Thouless, Anderson, Lieb and Palmer~\footnote{There
are other possible terms, for example $\sum_{i,k,l} J_{i,k}^{2} J_{k,l}^{2} J_{l,i}^{2}$, but they do vanish
in this limit.}.  The extra term can be omitted in the ferromagnetic case, because it gives a vanishing
contribution.

We must add a word of caution: the TAP free energy has not been derived from a variational equation: the
magnetization should satisfy the the stationary equation:
\be
{\partial F \over\partial m_{i}}=0 \ ,
\ee
but there is no warranty that the correct value of the magnetization  is the absolute minimum of the 
free energy. Indeed in many cases the paramagnetic solution (i.e. $m_{i}=0$) is the absolute minimum at all
temperatures, giving a value of the free energy density that would be equal to $-\frac14\beta$. The
corresponding internal
energy would be equal to $-\frac12 \beta$: it would give an result divergent at $T=0$, that in most of the
cases would be unphysical.

Let us be more specific (although these consideration are quite generic) and let us consider the
case of the Sherrington Kirkpatrick model, where all spins are connect and $z=N-1$.

Here one could be more precise and compute the corrections to the TAP free energy: an explicit computation
shows that these corrections are not defined in some regions of the magnetization space \cite{ANDERSON,DAT-P}.
When the corrections are not defined, the TAP free energy does not make sense.  In this way (after a
detailed computation) one arrives the conclusion that one must stay in the region where the following
condition is satisfied: 
\be \beta^{2} \mbox{Av}\left( (1-m_{i}^{2})^{2} \right) \le 1 \ , \label{DAT} \ee
where $\mbox{Av}$ stands for the average over all the points $i$ of the sample.  
When the previous relation is not valid, one finds that the correlations function, that were assumed to be
negligible, are divergent \cite{MPV} and the whole computation fails.
The correct recipe is to find the absolute minimum of the TAP free energy only in this region~\cite{ANDERSON}.
Of course the paramagnetic solution is excluded as soon as $T<1$.

Let us as look to the  precise expression of the TAP stationarity equations:
they are
\be
m_{i}=\tgh(\beta h^{eff}_{i})\ , \ \ \ \ \ h^{eff}_{i}=\sum_{k}
\left( J_{i,k}m_{k}-\beta m_{i}J_{i,k}^{2}(1-m_{k}^{2})\right) \ .
\ee
For large $N$ these equations can be simplified to
\be
h_{i}=\sum_{k}
\left( J_{i,k}m_{k}-\beta m_{i}(1-q_{EA})\right)\ ,
\ee
where 
\be
q_{EA}=\mbox{Av}(m_{i}^{2}) \ .
\ee

Apparently the TAP equations are more complex that the naive ones (\ref{NAIVE}); in reality their analysis is
simpler.  Indeed, using perturbation expansion with respect to the term $J_{i,k}$ that are of
order $O(N^{1/2})$, we can rewrite the effective field as
\be
h^{eff}_{i}=\sum_{k}  J_{i,k}m^{c}_{k} \ , \label{HTAP}
\ee
where the cavity magnetizations $m^{c}_{k}$ are the values of the magnetizations at the site $k$ in a system
with $N-1$ spins, where the spin at $i$ has been removed \cite{MPV} (the correct notation would be
$m^{c}_{k;i}$ in order to recall the dependence of the cavity magnetization on the spin that has removed, but
we suppress the second label $i$ when no ambiguities are present). The previous equations are also called
Bethe's equations because they were the starting point of the approximation Bethe in his study of the two
dimensional Ising model. 

The crucial step is based on the following relation:
\be
m_{k}=m^{c}_{k} +\beta m_{i} J_{i,k}(1- (m^{c}_{k;i})^{2}) \ .
\ee
This relation can be proved by using the fact that the local susceptibility $\partial m_{i} / \partial h_{i}$
is given by $\beta(1-m_{i}^2)$ in the mean field approximation.  Although the difference between $m(k)$ and
$m^{c}_{k}$ is small, i.e. O($N^{-1/2}$), one obtains that the final effect on $h_{eff}$ is of O(1).

The validity of eq.  (\ref{HTAP}) is rather fortunate, because in this formula $J_{i,k}$ and $m^{c}_{k}$ are
uncorrelated.  Therefore the central limit theorem implies  that, when one changes the $J_{i,k}$, the
quantity $h^{eff}_{i}$ has a Gaussian distribution with variance $q_{EA}$.  Therefore the probability distribution
of $h^{eff}$ is given
by
\be
P(h) dh= (2 \pi)^{-1/2} \exp(-q^2/2) dh \equiv d \mu_{q}(h) \label{PHNAIVE}
\ee

However this result is valid only under the hypothesis that there is a one to one correspondence of the
solutions of the TAP equation at $N$ and at $N-1$, a situation that would be extremely natural if the number
of solutions of the TAP equations would be a fixed number (e..g 3, as happens in the ferromagnetic case at low
temperature).  As we shall see, this may be not the case and this difference brings in all the difficulties and the
interesting features of the models.

\subsection{The cavity method: the naive approach}

Given the fact that all the points are equivalent, it is natural to impose the condition that the statistical 
properties of the spin at $i$, when the $J_{i,k}$ change, are $i$ independent.   For a large system 
this statistical properties coincide with those obtained by looking to the properties of the other
$N-1$ spins of the system.
This condition leads to the cavity method, where one compare the average properties of the spins in similar
systems with slightly different size.

The cavity method is a direct approach that in principle can be used to derive in an probabilistic way all the
results that have been obtained with the replica method~\footnote{The replica method is based on a saddle
point analysis of some $n$ dimensional integral in the limit where $n$ goes to zero.}.  The replica approach is
sometimes more compact and powerful, but it is less easy to justify, because the working hypothesis that are
usually done cannot be always translated in a transparent way in physical terms~\footnote{ Not all the results
of the replica approach have been actually derived using the cavity 
approach, although in principle it shuold be possible to do so.}.

The idea at the basis of the cavity method is simple: we consider a system with $N$ spins ($i=1,N$) and we
construct a new system with $N+1$ spins by adding an extra spin ($\sigma_{0}$).  We impose the consistency
condition that the average properties of the new spin (the average being done with respect to $J_{0,i}$) are
the same of that of the old spins \cite{MPV,MP1Be}.

In this way, if we assume that there is only one non-trivial solution to the TAP equations~\footnote{We
neglect the fact the if $m_{i}$ is a particular solution of the TAP equations, also $-m_{i}$ is a solution of
the TAP equations.  This degeneracy is removed if we add an infinitesimal magnetic field or we restrict
ourselves only the half space of configurations with positive magnetization.}, we get that for that particular
solution
\be
\overline{m_{0}^2}=\mbox{Av}(m_{i}^2) \ ,
\ee
where the bar denotes the average over all the $J_{0,i}$.

The Hamiltonian of the new spin is:
\begin{equation}
\sigma_{0}\sum_{i=1,N}J_{0,i}\sigma_{i} \ .
\end{equation}
If we suppose that the spins $\sigma_{i}$ have vanishing correlations and we exploit the fact
that each individual term of the Hamiltonian is small, we find that
\bea
m_{0}\equiv \lan \sigma_{0}\ran =\tanh (\beta h^{eff})\ . \\
h^{eff}=\sum_{i=1,N}J_{0,i}m^c_{i} \ ,
\eea
where $m^c_{i}$ (for $i\ne0$) denotes the magnetization of the spin $\sigma_{i}$ {\sl before} we add the 
spin 0.  In this way we have rederived the TAP equations for the magnetization in a direct way, under the
crucial
hypothesis of absence of long range correlation among the spins (a result that should be valid if we consider
the expectation values taken inside one equilibrium state). It should be clear why the whose approach does not
make sense when the condition eq. (\ref{DAT}) is not satisfied.

We have already seen that when the variables $J_{0,i}$ are random, the central limit theorem implies that $h$ is a Gaussian random variable with variance
\begin{equation}
\ba{h^{2}}= q_{EA}\equiv {\sum_{i=1,N}(m^c_{i})^{2}\over N} \ ,
\end{equation}
and the distribution probability of $h$ is given by eq. (\ref{PHNAIVE}). The same result would be true 
also when the variables $m^c_{i}$ are
random.

If we impose the condition that the average magnetization squared of the new point is
equal to average over the old points, we arrive to the consistency equation:
\begin{equation}
q_{EA}=\ba {m_{0}^{2}}= \int d\mu_{q_{EA}}(h)\tanh^{2} (\beta h)\label{QNAIVE} \ ,
\end{equation}
where $d\mu_{q_{EA}}(h)$ denotes a normalized Gaussian distribution with zero average and variance
$q_{EA}$.
It is easy to check \cite{MPV} that the $h$-dependent increase in the total free energy of the system (when we
go from $N$ to $N+1$) is
\begin{equation}
    \beta \Delta F(h) \equiv -\ln (\cosh(\beta h))  \ .
\end{equation}

It is appropriate to add a comment.  The computation we have presented relates the magnetization a spin of the
systems with $N+1$ spins to the magnetizations of the system with $N$ spins: they are not a closed set of
equations for a given system.  However we can also write the expression of magnetization at the new site as function
of the magnetizations of the system with $N+1$ spins, by computing the variation of the magnetizations in a
perturbative way.  Let us  denote by $m$ the magnetization of the old system
($N$ spins) and by $m'$ the magnetization of the new system ($N+1$ spins).
Perturbation theory tell us that
\be
m'_{i}\approx m_{i}+J_{0,i}m'_{0}{\partial m_{i}\over h_{i}}=
m_{i}+J_{0,i}m'_{0}\beta(1-(m'_{i})^{2}) \ .
\ee
It is not surprising that using the previous formula we get again the TAP equations \cite{TAP,MPV}:
\bea
m'_{0}=\tanh (\beta h) \,  \nonumber \\
h=\sum_{i=1,N}J_{0,i}m_{i} \approx \sum_{i=1,N}J_{0,i}m'_{i} -m'_{0} \sum_{i}J_{0,i}^{2}\beta (1-m'_{i})\, \\ 
\approx
\sum_{i=1,N}J_{0,i}m'_{i} -m'_{0} \beta (1-q_{EA}) \, \nonumber
\eea
where $(N+1)q_{EA})=\sum_{i=0,N}m'_{i}$  (we have used the relation  $\ba{J_{0,i}^{2}}=N^{-1}$).

A detailed computation  show that the free energy corresponding to a solution of the TAP equations is given by the 
TAP free energy.
The computation of the total free energy is slightly tricky. Indeed we must take care than if we just add one
spin the coordination number $z$ change and we have to renormalize the $J$'s. This gives an extra term that
should be taken into account.

The explicit computation of the free energy could be avoided by computing the internal energy density and
verifying a {\sl a posteriori} the guessed form of the free energy is correctly related to the internal
energy.  Here we can use a general argument: when the $J$'s are Gaussianly distributed the internal energy
density $e$ is given
\bea
e=\frac12\overline{\sum_{i}J_{0,i}\lan\sigma_{0}\sigma_{i}\ran}=
\frac1{2N}\overline{\sum_{i}\frac{\partial}{J_{0,i}} \lan\sigma_{0}\sigma_{i}\ran}=\\
\beta\frac1{2N}\overline{\sum_{i}(\lan(\sigma_{0}\sigma_{i})^2\ran -\lan\sigma_{0}\sigma_{i}\ran^2)}=
\frac12\beta(1-q^2) \ .
\eea
In the first step  we have integrated by part with respect to the Gaussian distribution of the $J$, in the
last step se have used the assumption that connect correlation are small and therefore
\be
\lan\sigma_{0}\sigma_{i}\ran^2\approx (m_{0}m_{i})^2\ .
\ee
 In both ways one obtain the expression for the TAP free energy, where the only parameter free is $q\equiv
 q_{EA}$; we can call this quantity $f(q)$; its explicit expression is 
 \be
 \beta f(q)=-\frac14\beta^2 (1-q^2)+\int d\mu_{q}(h) \ln\left(  \cosh(\beta h)\right)
 \ee
 
 The
 value of $q$ can be computed in two mathematical equivalent ways:
 \begin{itemize}
     \item We use the equation (\ref{QNAIVE}) to find a non-linear equation for $q$.
     \item We find the stationary point of the free energy respect to $q$ by solving the equation $\partial f
     /\partial q =0$.
 \end{itemize}
 An integration by part is the needed step to prove the equivalence 
 of the two computations.
 
 An explicit computation shows that the non-trivial stationary point (i.e $q\ne 0$), that should be relevant
 at $T<1$, is no not a minimum of the free energy,  as it  should be a maximum   for reasons that would be clear
 later.  In this way one finds the form of the function $q(T)$.  which turns out to be rather smooth:
 \begin{itemize}
     \item $q(T)$ is zero for $T>1$.
     \item $q(T)$ it is proportional to $1-T$ for $T$ slightly below 1.
     \item $q(T)$  behave as $1-A T$, with positive $A$ when $T$ goes to zero.
\end{itemize}
 
 When this solution was written down \cite{SK}, it was clear that something was wrong. The total entropy was negative at
 low temperature. This was particular weird, because in usual mean field models the entropy is just the
 average over the sites of the local entropy that is naturally positive. The {\sl villain} is the extra term
 that we have added to the naive free energy in order to get the TAP free energy:
 indeed one finds that the entropy density is given by
 \be
 S=\mbox{Av}(S(m_{i}))-\frac{(1-q)^2}{4T^2}\label{DISASTER} \ .
 \ee
 The average of the local term goes like $T$ at low temperatures, while the extra term goes to $-A^2/4$, i.e.
 -0.17. It is important to note that the extra term is needed to obtain that for $T>1$, $e=-\beta/2$, a
 results that can be easily checked by constructing the high temperature expansion.
 
 On the contrary numerical simulations \cite{SK} showed a non-negative entropy (the opposite would be an extraordinary
 and unexpected event) very similar to the analytic computation for $T>0.4$ and going to zero at $T^2$ (the
 analytic result was changing sign around $T=.0.3$).
 
 Also the value of the internal energy was wrong: the analytic computation gave $e =-0.798$ while the
 numerical computations gave $e=-0.765\pm 0.01$ (now numerical simulations give $e=-0.7633 \pm 0.0001)$.
 
It is clear that we are stacked, the only way out is to leave out the hypothesis of the existence of only one
solution of the TAP equations (or equivalently of only one equilibrium state) and to introduce this new
ingredient in the approach.  This will be done in the next section.

 \section{Many equilibrium states}
 \subsection{The definition of equilibrium states}
 The origin of the problem firstly became clear when De Almeida and Thouless noticed that in the whole region $T<1$, 
 the inequality (\ref{DAT}) was violated:
 \be
 \beta^{2} \mbox{Av}\left( (1-m_{i}^{2})^{2} \right) =1 + B_{0}(1-T)^2\label{AHI} \, ,
 \ee
with $B_{0}$ positive.

The problem became sharper when it was rigorously proved that the {\sl wrong} naive result was a necessary
consequence of the existence of a only one equilibrium state, i.e. of the assumption that connected correlations
functions are zero (in an infinitesimal magnetic field) and therefore
\be
\lan \sigma_{i} \sigma_{k}\ran \approx \lan \sigma_{i} \ran \lan \sigma_{k}\ran \ .
\ee

In order to develop a new formalism that overcomes the pitfalls of the naive replica approach, it is necessary
to discuss the physical meaning of these results and to consider the situation where many states are present.
The first step consists in introducing the concept of {\em pure states in a finite volume} \cite{MPV,CINQUE}.
This concept is crystal clear from a physical point of view, however it is difficult to state it in a
rigorous way.  We need to work at finite, but large volumes, because the infinite volume limit for local
observables is somewhat tricky and we have a chaotic dependence on the size \cite{NS,CINQUE}.

At this end we consider a system with a total of $N$ spins.  We partition the equilibrium configuration space
into regions, labeled by $\alpha$, and each region must carry a finite weight.  We define averages restricted to
these regions \cite{PAR1,PAR2,CINQUE}: {these regions will correspond to our finite volume pure states or
phases}.  It is clear that in order to produce something useful we have to impose sensible constraints on the
form of these partitions.  We require that the restricted averages on these regions of phase space are such
that connected correlation functions are small when the points are different in an infinite range model.  This
condition is equivalent to the statement that the fluctuation of intensive quantities \footnote{ Intensive
quantities are defined in general as $ b =\frac{1}{N} \sum_{i=1}^N B_i $ ,where the functions $B_i$ depend
only on the value of $\si_i$ or from the value of the nearby spins.} vanishes in the infinite volume limit
inside a given phase.

The phase decomposition depends on the temperature.  In a ferromagnet at low temperature  two regions are
defined by considering the sign of the total magnetization.  One region includes configurations with a
positive total magnetization, the second selects negative total magnetization.  There are ambiguities for
those configurations that have exactly zero total magnetization, but the probability that these configurations
are present at equilibrium is exponentially small at low temperature.

In order to present an interpretation of the results we assume that such decomposition exists also each
instance of our problem.  Therefore the {\em finite} volume Boltzmann-Gibbs measure can be decomposed in a sum
of such finite volume pure states and inside each state the connect correlations are small.  The states of the
system are labeled by $\al$: we can write

\begin{equation}
  \lan\  \cdot\  \ran =\sum_{\alpha} w_{\al}\lan\  \cdot\  
  \ran_{\al} \ ,
  \protect\label{E-WSUM}
\end{equation}
with the normalization condition

\begin{equation}
  \sum_{\alpha}w_{\alpha}=1 -\eps\ ,
  \protect\label{E-WNOR}
\end{equation}
where $\eps \to 0$ when $N \to \infty$.

The function $P_{J}(q)$ for a particular sample is given by

\begin{equation}
  P_{J}(q)=\sum_{\al,\beta} w_\al w_\beta \delta(q_{\al,\beta}-q)\ , \label{Q}
\end{equation}
where $q_{\al,\gamma}$ is the overlap among two generic configurations 
in the states $\alpha$ and $\gamma$:
\be
q_{\al,\gamma}={\sum_{1=1,N}m_{\alpha}(i)m_{\gamma}(i)\over N}\ ,
\ee
and $m_{\alpha}(i)$ and $m_{\gamma}(i)$ are  respectively the magnetizations within the state $\alpha$ and
$\gamma$.

For example, if we consider two copies (i.e. $\sigma$ and $\tau$) of the same system, we have;
\be
q[\sigma,\tau]={\sum_{1=1,N}\sigma(i)\tau(i)\over N}
\ee
and 
\be
\lan g(q[\sigma,\tau])\ran_{J}=\int  g(q) P_{J}(q) dq \ ,
\ee
where $g(q)$ is a generic function of $q$.

Given two spin configurations ($\si$ and $\tau$)  we can introduce the natural concept of 
distance by

\begin{equation}
  d^{2}(\si,\tau)\equiv \frac{1}{2N} \sum_{i=1}^N(\si_i-\tau_i)^2 \ ,
\end{equation}
that belongs to the interval [0 - 2], and is zero only if the two configurations are equal (it is one if they 
are uncorrelated).  In the thermodynamical 
limit, i.e. for $N\to\infty$, the distance of two configurations is zero if the number of different spins remains 
finite.  The percentage of different $\si$'s, not the absolute number, is relevant in this definition of the distance.  
It is also important to notice that at a given temperature $\beta^{-1}$, when $N$ goes to infinity, the number of 
 configurations inside a state is extremely large: it is proportional to $\exp (N { S}(\beta))$, where ${ S}(\beta)$ is 
the entropy density of the system). 

In a similar way we can introduce the distance between two states defined as the distance among two generic
configurations of the states:
\be
d^2_{\al,\gamma}=1-q_{\al,\gamma} \label{DIS}\ .
\ee
In the case where the self-overlap 
\be
q_{\alpha,\alpha}={\sum_{1=1,N}m_{\alpha}(i)^2\over N}\ ,
\ee
does not depend on $\alpha$ and it is denoted by $q_{EA}$, we have that the distance of a state with itself is
not zero:
\be
d^2_{\alpha,\alpha}=1-q_{EA} \ .
\ee

We expect that finite volume pure states will enjoy the following properties that hopefully characterizes the
finite volume pure states \cite{CINQUE,PAR1}:

\begin{itemize}

	\item When $N$ is large each state includes an exponentially large number of
	configurations~\footnote{We warn the reader that in the case of a glassy system it may be not possible to
	consider $N\to\infty$ limit of a given finite volume pure state: there could be no one to one
	correspondence among the states at $N$ and those at $2 N$ due to the chaotic dependence of the
	statistical expectation values with the size of the system \cite{NS}. }.
	
	\item The distance of two different generic configurations
	$C_{\alpha}$ and $C_{\gamma}$ (the  first belonging  to state $\alpha$
	and the second to state $\gamma$) does not depend on the
	$C_{\alpha}$ and $C_{\gamma}$, but only on $\alpha$ and
	$\beta$. The distance
	$d_{\alpha,\gamma}$ among the states $\alpha$ and $\gamma$, is
	the distance among two generic configurations in these two
	states. The reader has already  noticed that with this definition the
	distance of a state with itself is not zero. If we want, we can define an alternative
	distance:

	\begin{equation}
	  D_{\alpha,\gamma} \equiv
	  d_{\alpha,\gamma} -
	  \frac12\left(d_{\alpha,\alpha}+d_{\gamma,\gamma}\right)\ ,
	\end{equation}
	in such a way that the distance of a state with itself is zero
	($D_{\alpha,\alpha}=0$). 
	\item The distance between two configurations belonging to the
	same state $\alpha$ is strictly smaller than the distance
	between one configuration belonging to state $\alpha$ and a
	second configuration belonging to a different state
	$\gamma$. This last property can be written as

	\begin{equation} 
	  d_{\alpha,\alpha} < d_{\alpha,\gamma}\  \mbox{if}\ \alpha\ne\gamma\ .
	\end{equation} 
 This property forbids to have different states such that $D_{\alpha,\gamma}=0$, and it is crucial in avoiding
 the possibility of doing a too fine classification \footnote{For 
 example if in a ferromagnet (in a witless
 mood) we would classify the configurations into two states that we denote by $e$ and $o$, depending  if the
 total number of positive spins is even or odd, we would have that $d_{e,e}=d_{e,o}=d_{o,o}$.}.
 
	\item The classification into states is the finest one that satisfies the three former properties.

\end{itemize}

The first three conditions forbid a too fine classification, while the
last condition forbids a too coarse classification.

For a given class of systems the classification into states depends on the temperature of the system.  In some
case it can be rigorously proven that the classification into states is possible and unique
\cite{KASROB,RUELLE} (in these cases all the procedures we will discuss lead to the same result).  In
usual situations in Statistical Mechanics the classification in phases is not very rich.  For usual materials,
in the generic case, there is only one state.  In slightly more interesting cases there may be two states.
For example, if we consider the configurations of a large number of water molecules at zero degrees, we can
classify them as water or ice: here there are two states.  In slightly more complex cases, if we tune
carefully a few external parameters like the pressure or the magnetic field, we may have coexistence of three
or more phases (a tricritical or multicritical point).

In all these cases the classification is simple and the number of states is small.  On the contrary in the
mean field approach to glassy systems the number of states is very large (it goes to infinity with $N$), and a
very interesting nested classification of states is possible.  We note ``en passant'' that this behavior
implies that the Gibbs rule \footnote{The Gibbs rule states that in order to have coexistence of $K$ phases
($K$-critical point), we must tune $K$ parameters.  In spin glasses no parameters are tuned and the number of
coexisting phases is infinite!} is not valid for spin glasses.

We have already seen the only way out from the inconsistent result of the naive approach is to assume the
existence of many equilibrium states.  A more precise analysis will show that we need not only a few
equilibrium states but an infinite number and which are the properties of these states.

\subsection{The description of a many states system}
The next step consists in describing a system with many equilibrium state and to find out which is the most
appropriate way to code the relevant information,

Let us assume that a system has many states. In equation
\begin{equation}
  \lan\  \cdot\  \ran =\sum_{\alpha} w_{\al}\lan\  \cdot\  
  \ran_{\al} \ ,
  \protect\label{E-WSUM1}
\end{equation}
there is a large number of $w$'s that are non-zero.

How are we are supposed to describe such a system? First of all we should give the list of the $w_{\alpha}$.
In practice it is more convenient to introduce the free energy of a state $F_{\alpha}$ defined by 
\be
w_{\alpha}\propto \exp( -\beta F_{\alpha})\ ,
\ee
where the proportionality constant is fixed by the condition of normalization of the $w$'s, i.e.
eq. (\ref{E-WNOR}). Of course the free energies are fixed modulo an overall addictive constant,

The second problem it to describe the states themselves.  In this context  we are particularly interested in describing
the relations among the different states. At this end we can introduce different kinds of  overlaps:
\begin{itemize}
    \item The spin overlap, which we have already seen, i.e. 
    \be q_{\alpha,\gamma}=N^{-1}\sum_{i}\lan\sigma_{i}\ran_{\alpha}\lan\sigma_{i}\ran_{\gamma}\ . \ee
    \item The link (or energy) overlap, which is defined as: 
    \be
     q^L_{\alpha,\gamma}=N^{-1}\sum_{i,k}
     J_{i,k}\lan\sigma_{i}\sigma_{k}\ran_{\alpha}J_{i,k}\lan\sigma_{i}\sigma_{k}\ran_{\gamma}\ .\ee
     \item We could introduce also more fancy overlaps like 
     \be
     q^{(l)}_{\alpha,\gamma}=N^{-1}\sum_{i,k}
     J^{(l)}_{i,k}\lan\sigma_{i}\sigma_{k}\ran_{\alpha}J^{(l)}_{i,k}\lan\sigma_{i}\sigma_{k}\ran_{\gamma}\ ,
     \ee
     where $J^{(l)}$ denotes the $l$-th power of the matrix $J$.
     \item More generally, in the same way as the distances we can introduce overlaps that depend on the local
     expectation value of an operator $O(i)$
     \be
     q^{(O)}_{\alpha,\gamma}=N^{-1}\sum_{i}
     \lan O(i)\ran_{\alpha}\lan O(i)\ran_{\gamma}\ ,
     \ee

\end{itemize}
In principle the description of the system should be given by the list of all $F_{\alpha}$ and by the list of 
all possible overlaps a between all the pairs $\alpha,\gamma$. However it may happens, as it can be argued to 
happen in the SK model, that all overlap are function of the same overlap $q$. For example in the SK model we have
\be
q^L=q^2\ .
\ee
More complex formulae holds for the other overlaps.

In the case in short range model we may expect that
\be
q^L=l(q^2)\, 
\ee
where $l(q)$ is a model (and temperature) dependent function \cite{MaPa,CGGV}

If this {\sl reductio ad unum} happens, we say that we have {\sl overlap equivalence} and this is a property
that has  far reaching consequences.  The simplest way to state this properties is the following.  We
consider a system composed by two replicas ($\sigma$ and $\tau$) and let us denote by $\lan \cdot \ran_{q}$
the usual expectation Boltzmann-Gibbs expectation value with the constraint that
\be
q[\sigma,\tau]= q \ .
\ee
Overlap equivalence states that in this restricted ensemble of the connected correlation functions of the
overlap (and of the generalized overlaps) are negligible in the infinite range case (they would decay with the
distance in a short range model).

There is no \emph{a priori} compulsory need for assuming overlap equivalence; however it is by far the
simplest situation and before considering other more complex cases it is better to explore this simple
scenario, that it is likely to be valid at least in the SK model.

If overlap equivalence holds, in order to have a macroscopic description of the system, we must know only the
list of all $F_{\alpha}$ and the list of al $q_{\alpha,\gamma}$.  As we have already mentioned, we can also
assume that for large $N$ $q_{\alpha,\alpha}$ does not depend on $\alpha$ and it is equal to $q_{EA}$ (this
property is a consequence that the states cannot be distinguished one form the other using intrinsic
properties).  At the end the high
level description of the system is given by \cite{PAR1,V2}:
\be
\cD= \{ q_{EA}, \{F_{\alpha}\}, \{q_{\alpha,\gamma}\} \}\ ,
\ee
where the indices $\alpha$ and $\gamma$ goes from 1 to $\Omega(N)$, where $\Omega(N)$ is a function that goes 
to infinity with $N$. The precise way in which $\Omega(N)$ increases with $N$ is irrelevant because all high free energy
          states give a very small contribution to  statistical sums.
	  
In a random system the descriptor $\cal D$ is likely to depend on the system~\footnote{In a glassy non-random 
system it is quite possible that there is a strong dependence of $\cal D$ on the size $N$ of the system.},
therefore we cannot hope to compute it analytically. The only thing that we can do is to compute the probability
distribution of the descriptor $\cP(D)$. Some times one refer to this task as high level statistical mechanics
\cite{PAR1}
because one has to compute the probability distribution of the states of the system an not of the
configurations, as it is done in the old \emph{bona fide} statistical mechanics (as it was done in the low level
statistical mechanics).

\subsection{A variational principle}

Low level and high level statistical mechanics are clearly intertwined, as it can be see from the following
considerations, where one arrives to computing the free energy density of the SK model using a variational
principle.

Here we sketch only the main argument (the reader should read the original papers \cite{V1,V2}). Here for
simplicity we neglect the terms coming for the necessity of renormalize the $J$ when we change $N$; these
terms can be added at the price of making the formalism a little heavier.

Let us consider an SK system with $N$ particles with a descriptor $\cD_{N}$. If we add a a new site we obtain 
a descriptor $\cD_{N+1}$  where the overlaps do not change and the new free energies are given
by
\be
F_{\alpha}(N+1)=F_{\alpha}(N)+\Delta F_{\alpha}, \ \ \ \ \ \beta\Delta F_{\alpha}={-\ln (\cosh(\beta h_{\alpha})) }
\ee
where 
\be
h_{\alpha}=\sum_{i}J_{0,i}m_{\alpha}(i) \ .
\ee 

Also in the case where the $h_{\alpha}$ are Gaussian random uncorrelated, the fact that the new free energy
depends on $h_{\alpha}$ implies that at fixed value of the free energy $F_{\alpha}(N+1)$ the corresponding
value of the old free energy $F_{\alpha}(N)$ is correlated with $h_{\alpha}$.  It is a problem of conditioned
probability.  If we look at the distribution of $h_{\alpha}$, conditioned at the value of the old free energy
$F_{\alpha}(N)$, we find a Gaussian distribution, but if we look to the same quantity, conditioned at the value
of $F_{\alpha}(N+1)$, the distribution is no more Gaussian.  Therefore if we extract a state with a probability
proportional to $w_{\alpha}\propto \exp(- \beta F_{\alpha}(N+1)) $ the distribution of $h_{\alpha}$ will be not Gaussian.

Now it is clear that the $h_{\alpha}$ (conditioned to the old values of the free energy) are random Gaussian correlated variables because 
\be
\overline{h_{\alpha}h_{\gamma}}=q_{\alpha,\gamma}\ .
\ee
This correlation induces correlations among the $\Delta F_{\alpha}$.

Let us suppose for simplicity that we have succeeded in finding a choice of $\cP(\cD)$ that is
self-reproducing, i.e. a probability distribution such that
\be
\cP_{N+1}(\cD)=\cP_{N}(\cD) \ . \label{SELF}
\ee
We have neglect an overall shift $\Delta F(\cP)$ of the free energies, that obviously depend on $\cP$.
The self-reproducing property is  non-trivial as we shall see later. Notice than in going from $N$ to $N+1$ the 
$q$'s do not change, so the crucial point are the
correlations among the $q$'s and the $F$'s that may change going form $N$ to $N+1$.

A very interesting variational principle \cite{V2} tell us that the true free energy density $f$ is given
by
\be
f=\max_{\cP}\Delta F(\cP) \label{VAR} \ ,
\ee
which automatically implies that
\be
f\ge \Delta F(\cP) \ \ \ \forall \cP\ .
\ee

The reader should notice that there are no typos in the previous formulae. The free energy is the maximum,
and not the minimum, with respect to all possible probabilities of the descriptor. It is difficult to explain
why these inequalities are reversed with respect to the usual ones. 

We could mumble that in random system we have to do the average over the logarithm of the free energy and 
the logarithm has the opposite convexity of the exponential.  This change of sign is very natural also using
the replica approach, but it is not trivial to give a compulsory physical explanation.  The only suggestion we
can give is  {\sl follow the details of the proof}.

The mathematical proof is not complex, there is a simpler form of this inequality that has been obtained by
an interpolating procedure by Guerra \cite{V1} that is really beautiful and quite simple (it involves only
integration by part and the notion that a square in non-negative).  

We must add a remark: although the theorem tells us that the correct value of the free energy is obtained for a
precise choice of the probability $\cP^*(\cD)$ that maximize eq. (\ref{VAR}), there is no warranty that the actual
form of the  probability of the descriptors is given by $\cP^*(\cD)$. Indeed the theorem itself does not
imply that the concept of descriptor is mathematical sound;  all the heuristic interpretations of the
descriptor and of the meaning of the
theorem is done at risk of the physical reader.

While the lower bound is easy to use, for particular cases it 
coincides with Guerra's one \cite{V1}, in this formalism  it is
particularly difficult to find  a $\cP(\cD)$ such that it is the actual maximum.  The
only thing that we are able to do is to prove by other means \cite{TALA} which is the value of $f$ and to find
out a $\cP^*(\cD)$ that it saturate the bounds \cite{MPV}.  It uniqueness (if restricted to the class of
functions that satisfy eq. (\ref{SELF})), is far from being established.  It is clear that we miss some crucial
ingredient in order to arrive to a fully satisfactory situation.

\section{The explicit solution of the Sherrington Kirkpatrick model}
\subsection{Stochastic Stability}

The property of stochastic stability was introduced in a different context \cite{SS1,SS2,SS3}, however it is
interesting to discuss them in this perspective.  Stochastic stability stability is crucial because it strongly constrains
the space of all possible descriptors and it is the hidden responsible of most of the unexpected cancellations
or simplifications that are present in the computation of the explicit solution of the Sherrington Kirkpatrick
model.

Let us sketch the definition of stochastic stability in this framework.  Let us consider a distribution
$\cP(\cD)$ and let us consider a new distribution $\cP'(\cD)$ where
\be
F'_{\alpha}=F_{\alpha}+r_{\alpha} \ .
\ee
Here the $r$'s are random variables such that the correlation among $r_{\alpha}$ and $r_{\gamma}$ is a
function of only $q_{\alpha,\gamma}$.
We say that the distribution is stochastically stable iff the new distribution is equal to the previous one,
after averaging over the $r$'s, apart from an overall shift or the $F$'s, i.e. iff
\be
\cP(\cD)=\cP'(\cD) \ . \label{E-SS}
\ee

It is evident that, if a distribution is stochastically stable, it is automatically self-reproducing because the
the shifts in free energy can be considered random correlated variables.  This strongly suggest to consider
only  stochastic stable  $cP(\cD)$. Moreover 
general arguments, that where at the origin of the first version of stochastic stability
\cite{SS1,SS2,SS3}
 imply that $\cP(\cD)$ should be stochastically stable.

More precisely the original version of stochastic stability implies the existence of a large set of sum rules, 
which can be  derived in this context if we assume that the equation (\ref{E-SS}) is valid.
These sum rules give relations among the joint average of the overlap  of more {\em
real replicas} (i.e. a finite number of copies of the system in the
same realization of the quenched disorder).  One can show
\cite{CONTUCCI} that under very general assumptions of continuity that  also valid in
finite dimensional models for the appropriate quantities.

These relations connects the expectation values of products of various functions $P_{J}(q)$ to the expectation
value of the function $P_{J}(q)$ itself.

An example of these relations is the following
\bea
P(q)=\overline{P_{J}(q)}\ , \nonumber\\
P(q_{1,2},q_{3,4})\equiv \overline{P_{J}(q_{1,2})P_{J}(q_{3,4})}=\\
\frac23  P(q_{1,2})P(q_{3,4}) +
	  \frac13  P(q_{1,2})\delta(q_{1,2}-q_{3,4}) \ . \nonumber
\eea
where we have denoted by an bar the average over the $J$'s.	  

This relation is difficult to test numerically,  as far as in finite volume systems we do not have 
exact delta functions, therefore very often one test simplified versions of these equations.
For example  let  us consider four replicas of the system (all with the same Hamiltonian) and 
we  denote by
\begin{equation}
  E(\ldots)\equiv  \overline{\langle \ldots \rangle}
\end{equation}
the global average, taken both over the thermal noise and over the
quenched disorder. 
The previous equation implies that
\begin{equation}
	       E(q_{1,2}^{2} q_{3,4}^{2})    =
	  \frac23  E(q_{1,2}^{2})^{2} +
	  \frac13  E(q_{1,2}^{4})\ ,
\protect\label{G-A}
\end{equation}
This relation is very interesting: it can be rewritten (after some algebra), in a slightly different notation as
\be
\overline{\left(\lan A \ran_{J} -\overline{\lan A \ran_{J}}\right)^2}=
\frac13\overline{\lan \left(A -\lan A\ran_{J}\right)^2\ran_{J}} \ ,
\ee
with $A=q^2$. This relation implies that the size of sample to sample fluctuations of an observable (l.h.s.) is
quantitatively related to the average of the fluctuations of the same observable inside a given sample
(r.h.s.). This relation is very interesting because it is not satisfied in most of the non-stochastically
stable models and it is very well satisfied in three dimensional models \label{CINQUE}.

An other relation of the same kind is:
\begin{equation}
	       E(q_{1,2}^{2} q_{2,3}^{2})    =
	  \frac12  E(q_{1,2}^{2})^{2} +
	  \frac12  E(q_{1,2}^{4})\ .
\protect\label{G-B}
\end{equation}  
One could write an infinite set of similar, but more complex sum rules.

Stochastic stability is an highly non-trivial requirement (notice that  the union of two
stochastically stable systems is not stochastically stable). 

Therefore it is interesting to present here the
simplest non trivial case of stochastically stable system, where all the overlaps are equal
\be
q_{\alpha,\gamma}=q_{0}<q_{EA},   \ \ \ \ \forall \, \alpha,\gamma\ \label{ONESTEP0} \ .
\ee

In this case (which is usually called one-step replica-symmetry breaking)
we have only to specify the the weight of each state that (as usually) is given by
\begin{equation}
w_{\al}\propto \exp (-\beta F_{\al}).
\end{equation}
Let us consider the case where the joint probability of the $F$'s is such that the $F'$ are random independent
variable: the probability of finding an $F$ in the
interval $[F;F+dF]$ is given by
\be
\rho(F)dF \ .
\ee
The total number of states is infinite and therefore the function $\rho(F)$ has a divergent integral.
In this case it is easy to prove that stochastic stability implies that the function $\rho$ must be of the form
\begin{equation}
\rho( F)\propto \exp(\beta m  F)\ .\label{ONESTEP}
\end{equation}

At this end let us consider the effect of a perturbation of strength $\epsilon$ on the free energy of a state,
say $\alpha$.  The unperturbed value of the free energy is denoted by $F_\alpha$.  The new value of the free
energy $G_{\al}$ is given by $G_{\al}=F_{\al}+\eps r_{\al} $ where $ r_{\al}$ are identically distributed
uncorrelated random numbers.  Stochastic stability implies that the distribution $\rho(G)$ is the same as
$\rho(F)$.  Expanding to second order in $\epsilon$ we see that this condition leads to
\be
d\rho/dF\propto d^2\rho/dF^2 \ .
\ee
The only physical solution is given by eq.~(\ref{ONESTEP}) with an appropriate value of $m$.  The parameter
$m$ must satisfy the condition $m<1$, if the sum
\be
\sum_{\alpha}\exp(-\beta F_{\alpha})
\ee
is convergent.

We have seen that
stochastic stability fixes the form of the function $\rho$ and therefore connects in an inextricable way the
low and the high free energy part of the function $\rho$.

In this case an explicit computation show that the function $P(q)$ is given by
\begin{equation}
P(q)=m \delta(q-q_{0})+(1-m) \delta(q-q_{EA})\ .
\end{equation}

It is interesting that the same parameter $m$ enters both in the form of the function $\rho(F)$ at large
values of $F$ and in the form of the function $P(q)$ which is dominated by the lowest values of $F$, i.e.
those producing the largest $w$'s.  This result is deeply related to the existence of only one family of
stochastic stable systems with uncorrelated $F$'s.  

This relation is interesting from the physical point of view, because one could argue that the off-equilibrium
dynamics depends on the behavior of the function $\rho(\Delta F)$ at very large argument, and in principle it
could be not related to the static property that depend on the function $\rho$ for small values of the
argument.  However stochastic stability forces the function $\rho(\Delta F)$ to be of the form
(\ref{ONESTEP}), in all the range of $\Delta F$.  Consequently the high $F$ and low $F$ behavior are
physically intwined \cite{FMPP}.

Stochastic stable distribution have remarkable properties, an example of it can be found in the appendix I,
taken from \cite{MP1Be}.

Stochastic stability in finite dimensional system has many interesting applications. For example it is
 possible, by analyzing the properties of a single large system, to reconstruct the properties of the
 whole $\cP(\cD)$ averaged over the ensemble of systems to which the individual system naturally belong
 \cite{LOCO3,LOCO4}: in a similar way the fluctuation dissipation relations tells us that we can
 reconstruct the $P(q)$ function from the analysis of the response and the correlations during aging \cite{FMPP}.

\subsection{The first non trivial approximation}

We start by presenting here the first non trivial approximation, i.e. the so called one step replica symmetry
breaking. Let us assume that the probability distribution of the descriptor  is given by eqs. (\ref{ONESTEP0}) and
(\ref{ONESTEP}).  We need to compute both the shift in free energy and joint probability distribution of the
effective field and of the free energy.
 
The computation goes as follows.  We suppose that in the system with $N$ spins we have a population of states
whose total free energies $F_{\al}$ are distributed (when $F_{\al}$ is not far from a given reference value
$F^{*}$ that for lighten the notation we take equal to zero) as
\begin{equation}
\cN_{N}(F_{N}) \propto \exp (\beta m F_{N})\ .
\end{equation}
When we add the new spin, we will find a value of the field $h$ that depends on the 
state $\alpha$. We can now consider the joint probability distribution of the new free
energy and of the magnetic field.
We obtain
\begin{equation}
\cN_{N+1}(F,h)=\int dh P_{q_{EA}}(h) \int dF_{N} \cN_{N}(F_{N}) \delta(F-F_{N}-\Delta F(h)),
\end{equation}
where $P_{q_{EA}}(h)$ is the probability distribution of the effective magnetic field produced on the spin at 0,
for a generic state and it is still given by $d\mu_{q_{EA}}(h)$.  It is crucial to take into account that the
new free energy will differs from the old free energy by an energy shift that is $h$ dependent.  If we
integrate over $F_{N}$ and we use the explicit
exponential form for $\cN_{N}(F_{N})$ we find that
\bea
\cN_{N+1}(F,h)\propto \exp (\beta m F) \int d\mu_{q_{EA}}(h) \exp (\beta m \Delta F(h))) \\
\propto
\exp (\beta m F) P_{N+1}(h) \ .
\eea
The probability distribution of the field at \emph{fixed} value of the new free energy is given by
\begin{equation}
P_{N+1}(h) \propto \mu_{q_{EA}}(h) \exp (\beta m \Delta F(h)))=
\mu_{q_{EA}}(h) \cosh(\beta h)^{m} \ .
\end{equation}
It is clear that $P_{N+1}(h)$  different from $\mu_{q_{EA}}(h)$ as soon as $m\ne 0$. In this way 
we find the consistency equation of the replica approach for $q_{EA}$:
\be
q_{EA}={\int dh \mu_{q_{EA}}(h) \cosh(\beta h)^{m} \tanh(\beta h)^2
\over \int dh \mu_{q_{EA}}(h) \cosh(\beta h)^{m}}\ .
\ee

If we put everything together taking care of also the other terms coming for the renormalization of the $j's$
we find
\be
\beta f(q_{1},m)=-\beta^2(1-(1-m)q_{1}^2) +m^{-1}\ln\left(\int d\mu_{q_{1}}(h)\cosh(\beta h)^m\right),
\ee
where we have used the notation $q_{1}$ at the place of of $q_{EA}$. 

This result is obtained in the case where $q_{0}=0$. Otherwise we have
\bea
\beta f(q_{0},q_{1},m)=
\frac14\beta^2(1-(1-m)q_{1}^2-mq_{0}^2) +\\
m^{-1}\int d\mu_{q_{0}}h_{0}\ln\left(\int d\mu_{q_{1}-q_{0}}(h_{1})\cosh(\beta (h_{0}+h_{1})^m\right)\ .
\eea

The value of the parameter and of the free energy can be found by solving the equations
\be
{\partial f\over \partial q_{0}}={\partial f\over \partial q_{1}}={\partial f\over \partial m}\ .
\ee

A few comments are in order:
\begin{itemize}
    \item The probability distribution of $h$ at fixed value of the free energy of the $N$ 
    spins system ($P_{s}(h)$) is {\sl not} the probability distribution of $h$ at fixed value of the 
    free energy of the $N+1$ spins system $P_{N+1}(h)$: the two free energies differs by an $h$ 
    dependent addictive factor and they do not have a flat distribution (as soon as $m\ne 
    0$). The probability distribution of $h$ at a fixed value of the free energy of the $N$ 
    spins system is Gaussian, but the probability distribution of $h$ at fixed value of the  
    free energy of the $N+1$ spins system is not a Gaussian.
   
    \item Only in the case were $\cN_{N}(F_{N}) $ is an exponential distribution, $\cN_{N+1}(F,h)$ factorizes into the 
    product of an $F$ and an $h$ dependent factor and the $\cN_{N+1}(F)$ has the same form of $\cN_{N}(F)$.  
    Self-consistency can be reached only in the case of an exponential distribution for $\cN_{N}(F_{N})$. This
    fact is a crucial consequence of stochastic stability.
   
    \item The self consistency equations for the $q$ do not fix the value of $m$.  This is natural 
    because $m$ is an independent variational parameter and we must  maximize over $q_{0}$, $q_{1}$
    and $m$.  If one looks to the distribution of the metastable of having a free energy density higher that
    the ground state (problem that we cannot discuss here for reason of space \cite{LH2005,IRENE}) one finds
    that the distribution of the free energies inside the metastable states is characterized by a different
    value of $m$ than the ground state.  In this different context the equation $\partial f/ \partial m=0$
    determines which is the free energy of the metastable states that are true equilibrium states.
 \end{itemize}
 
If one analyze the solution of these equation one finds that the entropy is  still negative at low
temperature, but only by of a tiny amount and that the value of $B_{0}$ in eq. (\ref{AHI}) is smaller of
a factor 9 with respect to the previous computation. We are on the write track, but we have not reached our
goal.

\subsection{Ultrametricity}
Let us consider two generic spin configuration at equilibrium and let us use the label $c$ to denote it and
the label $\alpha$ to denote the state to which they belong.

Their  overlap satisfies the following properties (usually called one step replica symmetry breaking):
\begin{itemize} 
    \item If $c=c'$ the overlap  is equal to 1, 
    \item If $c\ne c'$, $\alpha=\alpha'$ the overlap is equal to $q_{1}$,
    \item If  $c\ne c'$, $\alpha\ne\alpha'$ the overlap is equal to $q_{0}$.
\end{itemize}

We see that we have the beginning of an hierarchical  scheme. It is clear that the simplest generalization
will  assume that states are grouped into clusters, that are labeled by $\gamma$ and the overlaps are given by
\begin{itemize} 
    \item If $c=c'$ the overlap is equal to 1, 
    \item If $c\ne c'$ , $\alpha=\alpha'$ the overlap is equal to $q_{2}$,
    \item If  $c\ne c'$, $\alpha\ne\alpha'$  and $\gamma=\gamma'$ the overlap is equal to $q_{1}$.
    \item If  $c\ne c'$, $\alpha\ne\alpha'$ and $\gamma\ne\gamma'$  the overlap is equal to $q_{0}$. 
\end{itemize}
This case is called two step 
replica symmetry breaking.

In order to specify the descriptor of the system we have to define the probability distribution of the weight 
$w$'s or equivalently of the free energy $F$'s.

The following ansatz is the only one compatible with stochastic stability \cite{PRzz}
\bea
F_{\alpha,\gamma}= F_{\alpha}+F_{\gamma}\, , \\
\rho(F_{\alpha})=\exp(\beta m_{2}F_{\alpha}) \, \ \ \ \ 
\rho(F_{\gamma})=\exp(\beta m_{1}F_{\gamma}) \, ,\nonumber
\eea
where we have \be
m_{1}<m_{2} \ . 
\ee
This condition implies that the number of states increases faster that number of clusters~\footnote{Sometimes one
may consider the case where this condition is violated: in this case one speaks of an inverted tree. The physical
meaning of this inverted tree is clarified in \cite{FV}; an inverted  tree for metastable states is discussed
in \cite{KPV}, however the physical implications are less clear. It seems to me that there are some hidden
properties that we do not fully understand.}.

If we compute the function $P(q)$ it turns out to be given
\be
m_1\delta(q-q_{0}) 
+(m_{2}-m_1)\delta(q-q_{1})+(1-m_1)\delta(q-q_{2})\ .
\ee
The free energy as a more complex expression, that can be written as a 
\bea
\beta f(q_{0},q_{1},q_{2},m_{1},m_{2})= \nonumber\\
\frac14\beta^2(1-(1-m_{2})q_{2}^2+(m_{2}-m_{1})q_{1}^2+m_{1}q_{0}^2) +\\
m_{1}^{-1}\int d\mu_{q_{0}}(h_{0})\ln\left(\int d\mu_{q_{1}-q_{0}}(h_{1})I(h_{0},h_{1})
^{m_{2}/m_{1}}\right)\ ,
\nonumber\\
I(h_{0},h_{1})= \int d\mu_{q_{2}-q_{1}}(h_{2}) \cosh(\beta (h_{0}+h_{1}+h_{2}))^{m_{2}} \ .\nonumber
\eea

At the end progress has been done, the value of $A_{0}$ in equation (\ref{AHI}) is much smaller and the zero
temperature entropy is nearly equal to zero (e.g. $S(0)=-.003)$. The results are nearly consistent but not
completely consistent. Moreover the zero temperature energy seems to be convergent: it was -.798 in the naive 
approach, -765 in the one step approach and -.7636 in the two step approach (the correct value is -.7633).

It is clear that we can go on introducing more steps of replica symmetry breaking:
at the third steps we
will to assume that clusters are grouped into families, that are labeled by $\delta$ and the overlaps are given by
\begin{itemize} 
    \item If $c=c'$ the overlap is equal to 1, 
    \item If $c\ne c'$ , $\alpha=\alpha'$ the overlap is equal to $q_{3}$,
    \item If  $c\ne c'$, $\alpha\ne\alpha'$  and $\gamma=\gamma'$ the overlap is equal to $q_2$.
    \item If  $c\ne c'$, $\alpha\ne\alpha'$, $\gamma\ne\gamma'$ and $\delta=\delta'$ the overlap is equal to $q_{1}$.
    \item If  $c\ne c'$, $\alpha\ne\alpha'$, $\gamma=\ne \gamma'$ and $\delta\ne\delta'$ the overlap is equal to $q_{0}$.
\end{itemize}
In the same way as before stochastic stability implies that 
\bea
F_{\alpha,\gamma,\delta}= F_{\alpha}+F_{\gamma}+F_{\delta}\, , \nonumber\\
\rho(F_{\alpha})=\exp(\beta m_{3}F_{\alpha}) \, ,\\
\rho(F_{\gamma})=\exp(\beta m_{2}F_{\gamma}) \, ,\nonumber \\
\rho(F_{\delta})=\exp(\beta m_{1}F_{\delta}) \, , 
\eea
where we have \be
m_{1}<m_{2}<m_{3} \ . 
\ee

In this case the function $P(q)$ is given by
\bea
m_1\delta(q-q_{0}) +(m_{2}-m_1)\delta(q-q_{1})+\\ 
(m_{3}-m_2)\delta(q-q_{2})+(1-m_3)\delta(q-q_{3})
\eea
This hierarchical construction of the probability distribution of the descriptor is called ultrametric, because
if we use the distance defined in eq.  (\ref{DIS}) the following inequality holds
\cite{MPV}:
\be
d_{\alpha,\beta}< \max(d_{\alpha,\gamma},d_{\gamma,\beta}) \ \ \forall \gamma \label{ULTRA}
\ee

If the the ultrametricity is satisfied, we can associate to the set of states a tree, formed by a root, by the
nodes of the the tree and by the leaves are the states.  States branching from the same node belongs to the
same cluster.  In the two step breaking all the nodes are the clusters.  In the three step breaking the nodes
are both the clusters and the families.  In order from above we have the root, the families, the clusters and the states.
In order to fix the probability distribution of the descriptor, we have to give the self distance of the
states, the distance of the states belonging to the same clusters, the distance of the states belonging to
different clusters, but to the same family, and the distance among the states belonging to different families
\cite{MPV,PARISICON,PARISITREE}.  As we have already remarked the probability distribution of the leaves is
fixed by stochastic stability,

At the present moment it is not rigorously know which of the following two alternatives are correct:
\begin{itemize}
    \item Ultrametricity is a necessary condition, i.e. a consequence of overlap
equivalence and of stochastic stability,
\item  Ultrametricity is an independent hypothesis. 
\end{itemize}
However there are strong arguments that
suggest that the first statement is true \cite{ULTRANEC} (although a rigorous proof is missing) and personally I
am convinced that this is the case.

The question if overlap equivalence (and consequently ultrametricity) is a consequence of stochastic stability
it is a more difficult question. Opposite arguments can be presented:
\begin{itemize}
    \item The fact that there is no known example of a stochastically stable system where we can define two
    non-equivalent overlaps, can be considered a strong suggestion of the non-existence of such systems.
    \item The construction of non-trivial stochastically stable system is rather difficult; in some sense it
    was done by chance using the algebraic replica method. It is likely that the construction of stochastically
    stable systems  with two non-equivalent overlaps will be rather complex (if it is non-void) and it may
    take a very strong effort to exhibit an explicit case.
\end{itemize}
It is clear that further research in this direction is needed: it would be very interesting to discover  a stochastically
    stable systems  with two non-equivalent overlaps. 
 
Generally speaking stochastic stability may have unexpected consequences: for example for a stochastically
stable system (in the original sense \cite{SS1,SS2,SS3}), without doing any assumption on the decomposition
into states, it is possible to prove \cite{PATA} that the support of the function $P(q)$ is discrete, i.e., if $N$ is
large, we can find $n$ very small intervals such that the integral of the function $P(q)$ outside these
intervals goes to zero with $n$, at least as $n^{-2}$.

 \subsection{Continuos Symmetry Breaking}

From the discussion of the previous section is completely evident that we can go on and introduce more and
more levels of symmetry breaking.  The problem is how to control the limit where the number of levels is going to
infinity.

At this end we can introduce 
the function $q(x)$ on the interval $[0:1]$ that is given by
\be
q(x)=q_{l} \ \  \mbox{if} \ \ \ m_{l}<x<m_{l+1}\ ,
\ee
where the index $l$ runs from zero to $K$ and we have used the convention that $m_{0}=0$ and $m_{K+1}=1$.
This function is interesting  for many reasons.
\begin{itemize}
    \item In the case of $K$-step symmetry breaking the function $q(x)$ is piecewise constant and takes $K+1$ 
    values.
    \item The inverse function $x(q)$ is discontinuous (we assume that the $q'$'s are increasing and we use the
    convention $x(0)=0)$. We have the very simple relation
   \be
   P(q)={d x\over d q} \ .
   \ee
   \item In the limit $K \to \infty$ the function $q(x)$ may become a continuos function and in this case we
   can speak of continuos replica symmetry breaking.
   \item The limit  $K \to \infty$ is rather smooth for any reasonable observable.
 \end{itemize}
 
From one side it is possible to construct in this way an infinite tree with branches at all the levels by a
painful explicit construction.  A priori it is not obvious that this construction has a well defined limit,
however it is possible to check the all the properties of the tree can be computed and explicit formulae can
be obtained \cite{MPV,PARISICON,PARISITREE,FRAPA1} for all the possible quantities.  It is possible to define
this tree in a rigorous directly in the continuos limit \cite{RUELLETREE} by stating the properties of
resulting measure.
 
 It crucial that although branching may happens at any level, we can reduce ourself to consider trees with
 only a finite number of branches.  Indeed  the number of equilibrium states is diverging when the
 volume goes to infinity, but most of the states carry a very small weight (i.e. they are high free energy
 states).  An explicit computation \cite{RUELLETREE,PARISITREE} shows that the total weight of the states with
 weight less than $\eps$  goes to zero when $\eps$ goes to zero, therefore removing an infinite number of
 states, that carry negligible weight, we remain with a finite number of them.  

 In this construction one finds the the only free quantity is the function $P(q)$ itself; its form 
 determines analytically the full 
 $
 \cP(\cD)
 $ \cite{PRzz}.

 \begin{figure}
     \includegraphics[width=0.80\textwidth]{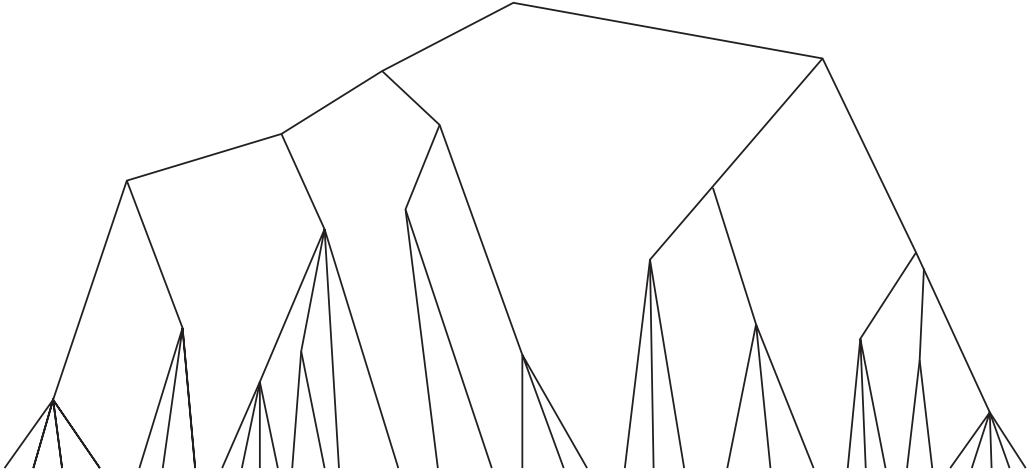}
     \caption{
     An example of a taxonomic tree of states in the case of an hierarchical breaking of the replica symmetry}
 \label{TASSI}
 \end{figure}

 The inequality (\ref{ULTRA}) implies that the most relevant states form an ultrametric tree, i.e.
 they can be put on a tree in such a way that  the states are on the branches of the tree and the distance
 among the states is the maximum level one has to cross for going from one states to an other.  In
 other words the states of a system can be classified in a taxonomic way.
 The most studies possibilities are the following:
 \begin{itemize}
	  \item All different states are at given distance $d$ one from the other and the taxonomy is
	  rather trivial (\emph{one step replica symmetry breaking}).
	  \item The states have a continuous distribution and branching points exist at any level
	  (\textit{continuous replica symmetry breaking}).  A suggestive drawing of how the tree may
	  look like, if we draw only the finite number of branches that have weight larger than $\eps$, is
	  shown in fig. (\ref{TASSI}).  A careful analysis of the properties of the tree can be found in
	  \cite{RUELLE,PARISITREE}.
  \end{itemize}
 Other distributions are possible, but they are less common or at least less studied.

 \subsection{The final result}
In order to compute the free energy is useful to write down a formula for its value that is valid also in the 
case where there is a finite number of steps. After some thinking, we can check that the following expression 
(\ref{CON})
is a compact way to write the previous formulae for the free energy in terms of the function $x(q)$ (or
$q(x)$).
The expression for the free energy is a functional (a functional of the function $q[x]$) and it is given by:
\be
\beta f[q(x)]={\beta^2 \over 4}\left(1- \int_0^1 q^2(x)\, dx\right)+f(0,0)\label {CON}
\ee
where the function $f(q,h)$ is defined in the strip $0\leq q \leq q_{EA}\equiv q_{K+1}$  and it obeys the non linear equation:
\be
\dot{f}=-(f''+x(q) (f')^2) \ ,
\label{eqf}
\ee
where dots and primes mean respectively derivatives with respect to $q$ and $h$. The initial condition is
\be
f(q,y)=  \cosh (\beta  m_{K}h) \ .
\ee
Now the variational principle tell us that the true free energy is greater than the maximum with respect to all
function $q(x)$ of the previous functional. It is clear that a continuos 
function may be obtained as limit of a stepwise continuos function and this shows how to obtain the result for 
a continuos tree in a smooth way

Talagrand \cite{TALA} is using on a cleaver the generalization of the bounds by Guerra \cite{V1} to systems
composed by a two identical copies: he finally arrives to write also an upper bound for the free energy and to
prove that they coincide.  In this way after 24 years the following formula for the free energy of the SK
model was proved to be true:
\be
f=\max_{q(x)}f[q(x)] \ . 
\ee
We have already remarked that the mathematical proof does not imply that the probability distribution of the
descriptors is the one predicted by the physical theory. The main ingredient missing is the proof that the
ultrametricity properties is satisfied. This is the crucial point, because ultrametricity and stochastic
stability (which is known to be valid) completely fix the distribution. It is likely that the needed step
would be to transform the estimates of \cite{ULTRABOUND} into rigorous bounds, however this is a rather
difficult task. Maybe we have to wait other 24 years for arriving to the rigorous conclusion that the heuristic approach 
gave the correct results, although I believe that there are no serious doubts on its full correctness.

 \section{Bethe lattices}
 \subsection{An intermezzo on random graphs}

 It is convenient to recall here the main  properties of random graphs \cite{Erdos_Renyi}. 

 There are many variants of random graphs: fixed local coordination number, Poisson distributed local
 coordination number, bipartite graphs\ldots They have the same main topological structures in the
 limit where the number ($N$) of nodes goes to infinity.

 We start by defining the random Poisson graph in the following way: given $N$ nodes we consider the
 ensemble of \emph{all} possible graphs with $M=\alpha N$ edges (or links).  A random Poisson graph is a
 generic element of this ensemble.
    
 The first quantity we can consider for a given graph is the local coordination number $z_{i}$, i.e.
 the number of nodes that are connected to the node $i$.  The average coordination number $z$ is the
 average over the graph of the $z_{i}$:
 \be
 z=\frac{\sum_{i=1,N}z_{i}}{N}\ .
 \ee
 In this case it is evident that
 \be
 z=2 \alpha \ .
 \ee
 It takes a little more work to show that in the thermodynamic limit ($N \to \infty$), the probability 
 distribution of the local coordination number is a Poisson distribution with average $z$.

 In a similar construction two random points $i$ and $k$  are connected with a
 probability that is equal to $z/(N-1)$. Here it is trivial to show that the probability
 distribution of the $z_{i}$ is Poisson, with average $z$. The total number of links is just
 $zN/2$, apart  from corrections proportional to $\sqrt{N}$. The two Poisson ensembles, i.e. fixed
 total number of links and fluctuating total number of links, cannot be distinguished locally for
 large $N$ and most of the properties are  the same.

 Random lattices with fixed coordination number $z$ can be easily defined; the ensemble is just given by
 all the graphs with $z_{i}=z$ and a random graph is just a generic element of this ensemble .

 One of the most important facts about these graphs is that they are locally a tree, i.e. they are
 locally cycleless.  In other words, if we take a generic point $i$ and we consider the subgraph
 composed by those points that are at a distance less than $d$ on the graph \footnote{The distance
 between two nodes $i$ and $k$ is the minimum number of links that we have to traverse in going from
 $i$ to $k$.}, this subgraph is a tree with probability one when $N$ goes to infinity at fixed $d$.
 For finite $N$ this probability is very near to 1 as soon as
 \be
 \ln (N) > A(z)\; d \ ,
 \ee
 $A(z)$ being an appropriate
 function. For large $N$ this probability is given by $1-O(1/N)$.

 If $z>1$ the nodes percolate and  a finite fraction of the graph belongs to a single giant connected
 component.  Cycles (or loops) do exist on this graph, but they have typically a length proportional
 to $\ln(N)$.  Also the diameter of the graph, i.e. the maximum distance between  two points of the
 same connected component is proportional to $\ln(N)$.  The absence of small loops is crucial because
 we can study the problem locally on a tree and we have eventually to take care of the large loops
 (that cannot be seen locally) in a self-consistent way.  i.e. as a boundary conditions at infinity.
 
 Before applying the problem will be studied explicitly in the next section for the ferromagnetic Ising model.
 \subsection{The Bethe Approximation in D=2}

 Random graphs are sometimes called Bethe lattices, because a spin model on such a graph can be
 solved exactly using the Bethe approximation.
 Let us recall the Bethe approximation for the two dimensional Ising model. 

 In the standard mean field approximation, one writes a variational principle assuming the all the
 spins are not correlated \cite{I}; at the end of the computational one finds that the magnetization satisfies
 the well known equation
 \be
 m=\tgh( \beta J z m) \ ,
 \ee
 where $z=4$ on a square lattice ($z=2d$ in $d$ dimensions) and $J$ is the spin coupling ($J>0$ for a
 ferromagnetic model).  This well studied equation predicts that the critical point (i.e. the point
 where the magnetization vanishes) is $\beta_{c} =1/z$.  This result is not very exiting in two
 dimensions (where $\beta_{c}\approx .44$)  and it is very bad in one dimensions (where
 $\beta_{c}=\infty$).  On the other end it becomes more and more correct when $d \to \infty$.

 A  better approximation can be obtained if we look to the system locally and we compute the
 magnetization of a given spin ($\sigma$) as function of the magnetization of the nearby spins
 ($\tau_{i}$,  $i=1,4$). If we assume that the spins $\tau$ are uncorrelated, but have magnetization
 $m$, we obtain that the magnetization of the spin $\sigma$ (let us call it $m_{0}$) is given by:
 \be
 m_{0}= \sum_{\tau} P_{m}[\tau] \tgh(\beta J \sum_{i=1,4}\tau_{i}) \ ,
 \ee
 where 
 \be
 P_{m}[\tau]= \prod_{i=1,4} P_{m}(\tau_{i}), \ \ \ \ \ \  
 P_{m}(\tau)=\frac{1+m}{2}\delta_{\tau,1}+\frac{1-m}{2}\delta_{\tau,-1} \ .
 \ee
 The sum over all the $2^{4}$ possible values of the $\tau$ can be easily done. 

 If we impose the
 self-consistent condition
 \be
 m_{0}(m)=m \ ,
 \ee
 we find an equation that enables us to compute the value of the magnetization $m$.

 This approximation remains unnamed (as far as I know) because with a little more work we can get
 the better and simpler Bethe approximation. The drawback of the previous approximation is that the 
 spins $\tau$ cannot be uncorrelated because they interact with the same spin $\sigma$: the effect
 of this correlation can be taken into account ant this leads to the Bethe approximation.

 Let us consider the system where the spin $\sigma$ has been removed. There is a cavity in the
 system and the spins $\tau$ are on the border of this cavity. We assume that in this situation these 
 spins are uncorrelated and  they have a magnetization $m_{C}$. 
 When we add the spin $\sigma$,
 we find that the probability distribution of this spin is proportional to
 \be
 \sum_{\tau} P_{m_{C}}[\tau]) \exp\left(\beta J \sigma \sum_{i=1,4}\tau_{i}\right) \ .
 \ee
 The magnetization of the spin $\sigma$ can be computed and after some simple algebra we get
 \be
 m=\tgh\{z \;\arth[ \tgh(\beta J) m_{C}] \}\ , \label{BETHE}
 \ee
 with $z=4$.

 This seems to be a minor progress because we do not know $m_{C}$. However we are very near the
 final result. We can remove one of the spin $\tau_{i}$ and form a larger cavity (two spins removed).
 If in the same vein we assume that the spins on the border of the cavity are uncorrelated and they
 have the same magnetization $m_{C}$, we obtain
 \be
 m_{C}=\tgh\{(z-1) \arth[ \tgh(\beta J) m_{C}] \}\ . \label{CAVITY}
 \ee
 Solving this last equation we can find the value of $m_{C}$ and using the previous equation we can 
 find the value of $m$.
		     
 It is rather satisfactory that in 1 dimensions ($z=2$) the cavity equations become
 \be
 m_{C}=\tgh(\beta J) m_{C}\ .
 \ee
 This equation for finite $\beta$ has no non-zero solutions, as it should be.

 The internal energy can be computed in a similar way: the energy density per link is
 given by
 \be
 E_{link}={ \tgh(\beta J) +m_{C}^{2}\over 1+ \tgh(\beta J)m_{C}^{2}}\ .
 \ee
We can obtain the free energy by integrating the internal energy as function of $\beta$.

 In a more sophisticated treatment we write the free energy as function of $m_{C}$:
 \be
 {\beta F(m_{C}) \over N}= F_{site}(m_{C})- \frac{z}{2} F_{link}(m_{C}) \ , \\
 \ee
 where $F_{link}(m_{C})$ and 
 $F_{site}(m_{C})$ are appropriate functions \cite{HH}.
 This free energy is variational, in other words the equation
 \be
 {\partial F \over \partial m_{C}} =0
 \ee
 coincides with the cavity equation (\ref{CAVITY}).
 
 \subsection{Bethe lattices and replica symmetry breaking}

 It should be now clear why the Bethe approximation is correct for random lattices. 
 If we remove a
 node of a random lattice, the nearby nodes (that were at distance 2 before) are now at a very large
 distance, i.e. $O(\ln(N))$. In this case we can write
 \be
 \lan \tau_{i_{1}}\tau_{i_{2}} \ran \approx m_{i_{1}}m_{i_{2}}
 \ee
 and everything seems easy. 

 This is actually easy in the ferromagnetic case where in absence of magnetic field at low temperature
 the magnetization may take only two values ($\pm m$). In more complex cases, (e.g.
 antiferromagnets) there are many different possible values of the magnetization because there are
 many equilibrium states and everything become complex (as it should) because the cavity equations
 become equations for the probability distribution of the magnetizations \cite{MPV}. 

 I will derive the TAP cavity equations on a random Bethe lattice \cite{TAP,MP1Be,FL,white} where the average number
 of neighbors is equal to $z$ and each point has a Poisson distribution of neighbors. In the limit where $z$
 goes to infinity we recover the SK model.

 Let us consider a node $i$; we denote by $\partial i$ the set
 of nodes that are connected to the point $i$. 
 With this notation the Hamiltonian can be written
 \be
 H=\frac12 \sum_{i}\sum_{k\in \partial i}J_{i,k}\sigma(i)\sigma(k)\, .
 \ee

 We suppose that for a given value of $N$ the system is in a state labeled by $\alpha$ and we suppose such a
 state exists also for the system of $N-1$ spins when the spin $i$ is removed.  Let us call $m(i)_{\alpha}$ the
 magnetization of the spin $k$ and $m(k;i)_{\alpha}$ the magnetization of the spin $k$ when the site $i$ is
 removed.  Two generic spins are, with probability one, far on the lattice: they are at an average distance of
 order $\ln(N)$; it is reasonable to assume that in a given state the correlations of two generic spins are
 small (a crucial characteristic of a state is the absence of infinite-range correlations).  Therefore the
 probability distribution of two generic spins is factorized (apart from corrections vanishing in probability
 when $N$ goes to infinity) and it can be written in terms of their magnetizations.

 We have already seen that the usual strategy is to write the equation directly for the cavity
 magnetizations. We obtain (for $l\in\partial i$):
 \be
 m(i;l)=\tanh \left( \sum_{k\in \partial i; k\ne l} \artanh(\tanh (\beta J_{i,k})m(k;i)) \right) \ . \label{MAG}
 \ee 
 Following this strategy we remain with $Nz$ equations for the $Nz$ cavity magnetizations $m(i;k)$; the true
 magnetizations ($m(i)$) can be computed at the end using equation (\ref{MAG}).

 In the SK limit (i.e. $z\approx N $), it is convenient  to take  advantage of the fact that the 
 $J$'s are proportional to
 $N^{-1/2}$ and therefore the previous formulae may be partially linearized; using the the law of  large numbers
 and the central limit theorem one recovers the previous result for the SK model.
 
 It is the cavity method, where one connects the magnetizations
 for a system of $N$ spins with the magnetizations for a system with $N-1$ spins.  In absence of the
 spin $i$, the spins $\sigma(k)$ ($k \in \partial i$) are independent from the couplings $J_{i,k}$.
 
 We can now proceed exactly as in SK model. We face  a new difficulty already at the replica symmetric level:
 the effective field
 \be
 h_{eff}=\sum_{k\in \partial i; k\ne l} \artanh\left(\tanh (\beta J_{i,k})m(k;i)\right)
 \ee
 is the sum of a finite number of terms and it no more Gaussian also for Gaussian $J$'s. This implies that the 
 distribution probability of $h_{eff}$ depends on the full probability distribution of the magnetizations and
 not only on its second moment. If one writes down the equations, one finds a functional equation for the
 probability distribution of the magnetizations ($P(m)$).
 
 When replica symmetry is broken, also at one step level, in each site we have a probability distribution of
 the magnetization over the different states $Q(m)$ and the description of the probability distribution of the
 magnetization is now given by the probability distribution $P[Q(m)]$, that is a functional of the
 probabilities $Q(m)$ \cite{MP1Be,MP2Be,white,MPZ,PARS}. 
 
 The computation becomes much more involved, but at least in the case of one step replica breaking (an
 sometimes at the two steps level) they can be carried up to the end. The one step approximation is quite
 interesting in this context, because there are optimization models in which it gives the exact results
 \cite{MPZ,MM}. A complete discussion of the stability of the one step replica approach can be found in
 \cite{TRE}.

 \section{Finite dimensions}
 \subsection{General considerations}
 It is well known that mean field theory is correct only in the infinite dimensional limit. Usually in high
 dimensions it gives the correct results. Below the upper critical dimensions (that is 6 in the case of spin
 glasses) the critical exponents at the
 transition point change and they can be  computed near to the upper critical dimensions using the
 renormalization group.
 
 Sometimes the renormalization group is needed to study the behavior in the low temperature phase (e.g. in the
 $O(n)$ spin models), if long range correlations are present.  A careful and complete analysis of spin glasses
 in the low temperature phase is still missing due to the extreme complexity of the corrections to the mean
 field approximation.  Many very interesting results have been obtained \cite{LOOP}, but the situation is
 still not clear.  May be one needs a new approach, e.g starting to compute the corrections to mean field
 theory at zero temperature.
 
 There are two different problems that we would like to understand much better:
 \begin{itemize}
     \item The value of the lower critical dimension.
     \item If the predictions of the mean field theory at least approximately are satisfied in three dimensions
  \end{itemize}
  
 These two problems will be addressed in the next two sections
 
 \subsection{The lower critical dimension}
 
 The calculation of free energy increase  due to an interfaces is a
 well known method to compute 
 the lower critical dimension in the case of  spontaneous symmetry breaking. I will recall the basic points
 and I will try to apply it to spin glasses.

 Generally speaking in the simplest case we can
 consider a system with two possible coexisting phases ($A$ and $B$), with different values of the order
 parameter,

 For standard ferromagnets  we may have
 ;
 \begin{itemize}
 \item { A: spins up.}
 \item B: spins down.
 \end{itemize}

 We will study what happens in a finite system in dimensions $D$ of size
 $M^d \  L$ with  $d=D-1$. 

 We put the system in phase $A$ at
 $z=0$ and in phase $B$ at  $z=L$. 
 The free energy of the interface is the
 increase in free energy due to this choice of boundary conditions with
 respect to choosing the same phase at $z=0$ and $z=L$.
 In many cases we have that the free energy increase $\delta F(M,L)$
 behaves for large $M$ and $L$ as:

 \be
 \delta F(L,M)= M^d/L^\omega \ ,
 \ee
 where $\omega$ is independent from the dimension.

 There is a  lower critical
 dimension where the free energy of the interface is finite:
  \be
 D_{c}=\omega+1 \, ,
 \ee
 and
 \be
 \delta F(L,L)= L^{(D-D_{c})}\ .
 \ee

 Heuristic arguments, which sometimes can be made rigorous, tell us that when $D=\omega+1$, (the lowest
 critical dimension) the two phases mix in such a way that symmetry is restored.

In most cases the value of $\omega$ from mean field theory is the exact one and therefore we can calculate in
this way the value of the lower critical dimension.  The simplest examples are the ferromagnetic Ising model
$\omega=0$ and the ferromagnetic Heisenberg model $\omega=1$.

For spin glasses the order parameter is the overlap $q$
 and all values of $q$ in the interval $[-q_{EA},q_{EA}]$ are allowed.
 We
 consider two replicas of the same system described by a Hamiltonian:
 \be
 H=H[\sigma]+H[\tau]\, ,
 \ee
 where $H$ is the Hamiltonian of a a single spin glass.
 We want to compute the free energy increase corresponding
 to imposing an expectation value of $q$ equal to $q_1$ at
 $z=0$ and $q_2$ at $z=L$.

 A complex computation gives \cite{INTERFACE} (for small $|q_1-q_2|$)
 \be
 \delta F \propto M L (|q_1-q_2|/L)^{5/2} \propto L^{(D-5/2)} \ .
 \ee
As a consequence, the {\sl naive}
 prediction of mean field theory for the lower critical dimension for
 spontaneous replica symmetry breaking is $D_c=2.5$.
 I stress that these predictions are {\sl naive}; corrections to the mean
 field theory are neglected. In known cases these kind of computations give the correct result,

 A direct check of this prediction has never been done, however it quite remarkable that there are numerical
 results that strongly suggest that the lower critical dimension is really 2.5. Indeed a first method based on
 finding the dimension where the transition temperature is zero \cite{OLD,Boettcher}, gives $D_c=2.491$. A more
 accurate method based on interpolating the value of a critical exponent \cite{Boettcher} gives
 \be
 D_c=2.4986\ ,
 \ee
 with an error of order $10^{-3}$ (my estimate).
 
 It seem that this prediction for the lower critical dimension is well satisfied. It success implies that  
 tree dimensional systems should be describe in the low temperature phase by the mean field approximation,
 although the correction could slowly vanish with the system size and  the critical exponents could be quite
 different from the naive one, due to the fact three dimensions are only 0.5 dimensions above the lower critical
 dimensions. What happens in three dimension will be discussed in the next section.

\subsection{The three dimensional case}

Let us describe some of the equilibrium properties that has be  computed or measured in the there dimensional 
case. We have to make a very drastic selection of the very vast literature, for a review see \cite{CINQUE,NUM1}.

\begin{figure}
  \includegraphics[width=0.33\columnwidth,angle=270]{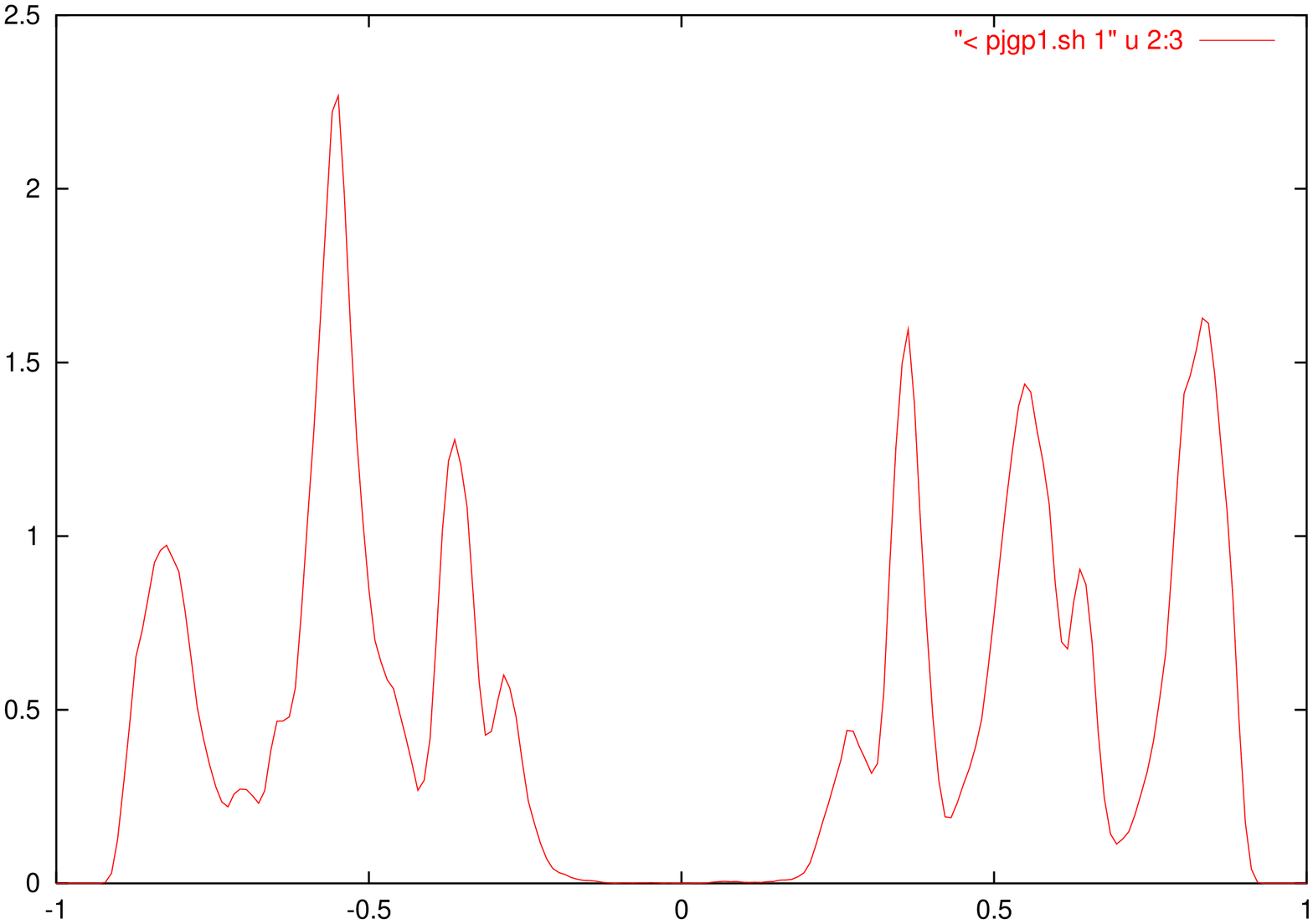}
  \includegraphics[width=0.33\columnwidth,angle=270]{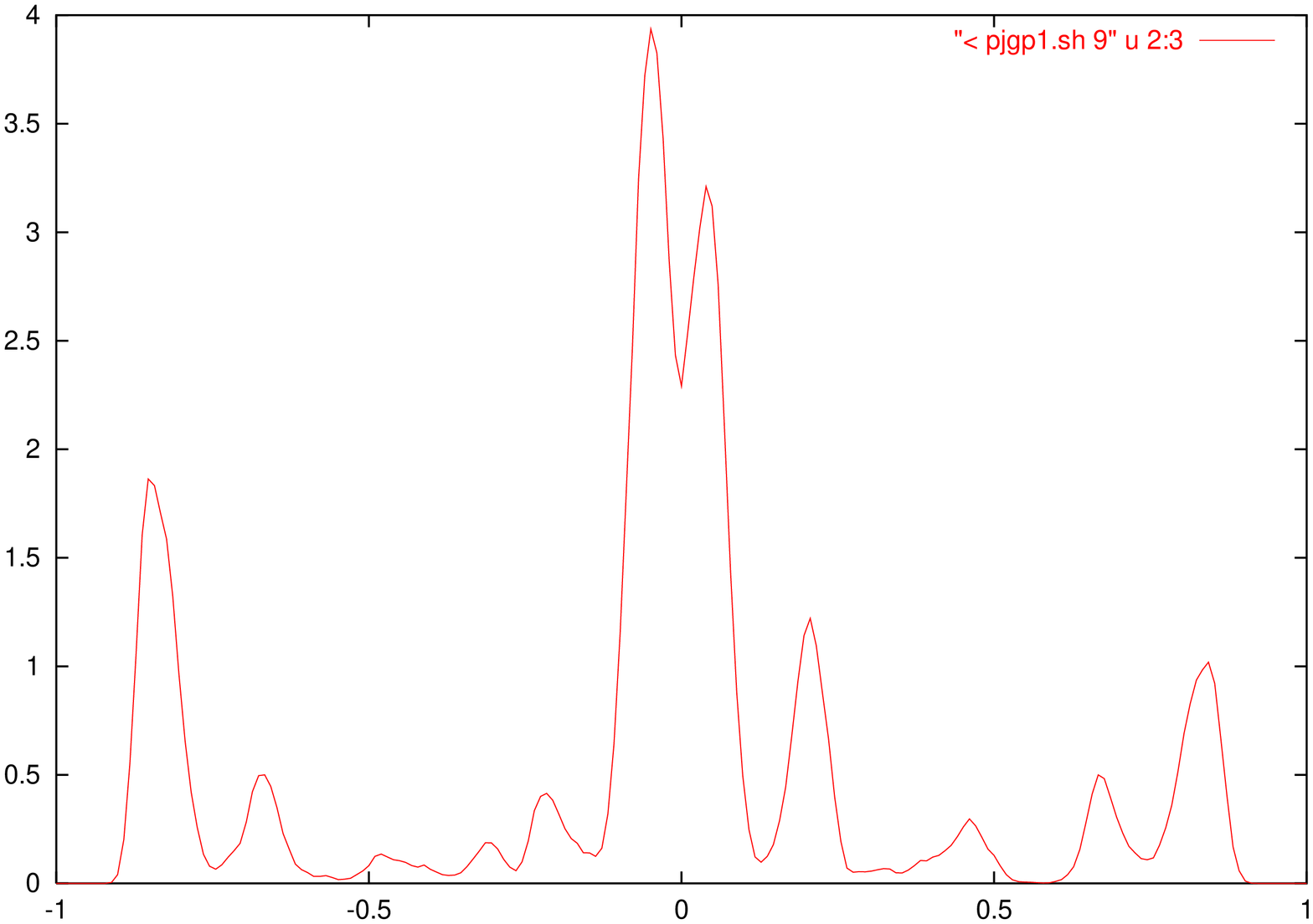}
  
  \caption{The function $P_{J}(q)$ for two  
  samples (i.e two choices of $J$) for $D=3,\  L=16$ ($16^{3}$ spins) from \cite{NUM1}.}
\label{A}
\end{figure}

All the simulations done with systems up $24^3$ \cite{GPR} find that at low temperature at equilibrium a
large spin glass system remains for an exponentially large time in a small region of phase space,
but it may jump occasionally in a relatively short time to an other region of phase space (it is like the
theory of {\it punctuated equilibria}: long periods of \emph{stasis}, punctuated by fast changes).  Example of
the function $P_{J}(q)$, i.e. the probability distribution of the overlap among two equilibrium
configurations, is shown in fig.  (\ref{A}) for two three dimensional 
  samples.

\begin{figure}
  \includegraphics[width=.70\textwidth]{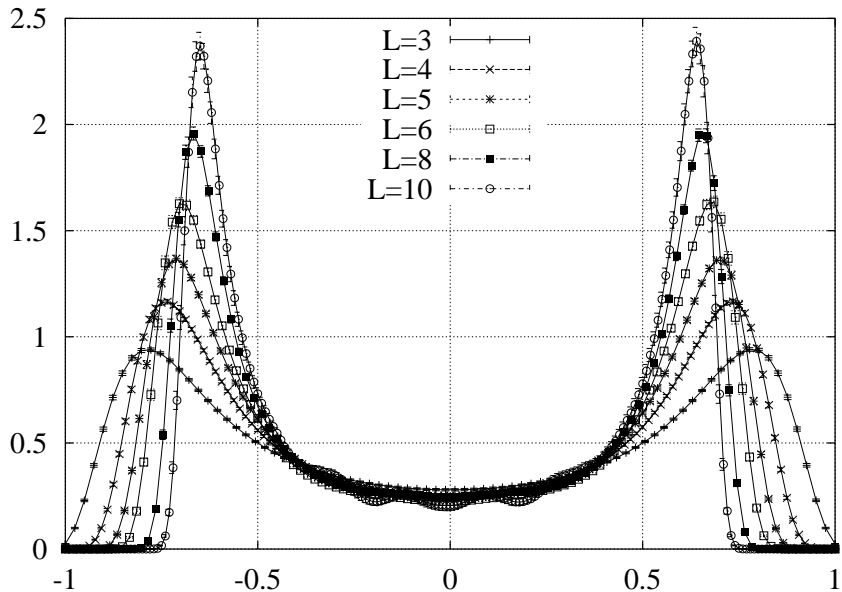}
  
  \caption{The function $P(q)=\ba{P_{J}(q)}$ after average over many samples (L=3\ldots 10) from \cite{NUM2}. }
\label{AVE}
\end{figure}

We have defined  $ P(q) \equiv \overline {P_{J}(q)} $, where the average is done over the different
choices of the couplings $J$, see fig.  (\ref{AVE}).  We have seen that average is needed because the theory
predicts (and numerical simulations also in three dimensions do confirm) that the function
$P_{J}(q)$ changes dramatically from system to system. It is clear that the function remains non-trivial in
the infinite volume limit.

In the mean field approximation the function $P(q)$ (and its fluctuations from system to system) can
be computed analytically together with the free energy: at zero magnetic field $P(q)$ has two delta
functions at $\pm q_{EA}$, with a flat part in between. The shape of the three dimensional function $P(q)$ is 
not very different from the one of the mean fields models.

The validity of the sum rules derived from stochastic stability (eqs. (\ref{G-A},\ref{G-B})) has been rather carefully verified
\cite{CINQUE}. This is an important test of the theory because they are derived in the infinite volume limit
and there is no reason whatsoever that they should be valid in a finite system, if the system would be not
sufficient large to mimic the behavior of the infinite volume system.  There are also indication that both
overlap equivalence \cite{MaPa,CGGV} and ultrametricity \cite{CMP,ULTRANEW} are satisfied, with corrections
that goes to zero slowly by increasing the system size.

Very interesting phenomena happen when we add a very small magnetic field. They are very important because the
magnetic properties of spin glass can be very well studied in real experiment.

From the theoretical point of view we expect that order of the states in free energy is scrambled when we
change the magnetic field \cite{MPV}: their free energies differ of a factor $O(1)$ and the
perturbation is of order $N$. Different results should be obtained if we use different experimental protocols:
\begin{itemize}
    \item
If we add the field at low temperature, the system remains for a very large time in the same state, only
asymptotically it  jumps to one of the lower equilibrium states of the new Hamiltonian.
\item
If we cool the system from high temperature in a field, we likely go directly to one of the 
good lowest free energy states.
\end{itemize}
 
Correspondingly there are two susceptibilities that can be measured also experimentally:
\begin{itemize}
    \item
The so called linear response susceptibility $\chi_{LR}$, i.e. the response within a state, that is
observable when we change the magnetic field at fixed temperature and we do not wait too much.  This
susceptibility is related to the fluctuations of the magnetization inside a given state.
\item
The true equilibrium susceptibility, $\chi_{eq}$, that is related to the fluctuation of the
magnetization when we consider also the contributions that arise from the fact that the total
magnetization is slightly different (of a quantity proportional to $\sqrt{N}$) in different states.
This susceptibility  is very near to $\chi_{FC} $, the field cooled susceptibility, where one
cools the system in presence of a field.
\end{itemize}

\begin{figure}
\includegraphics[width=.49\columnwidth]{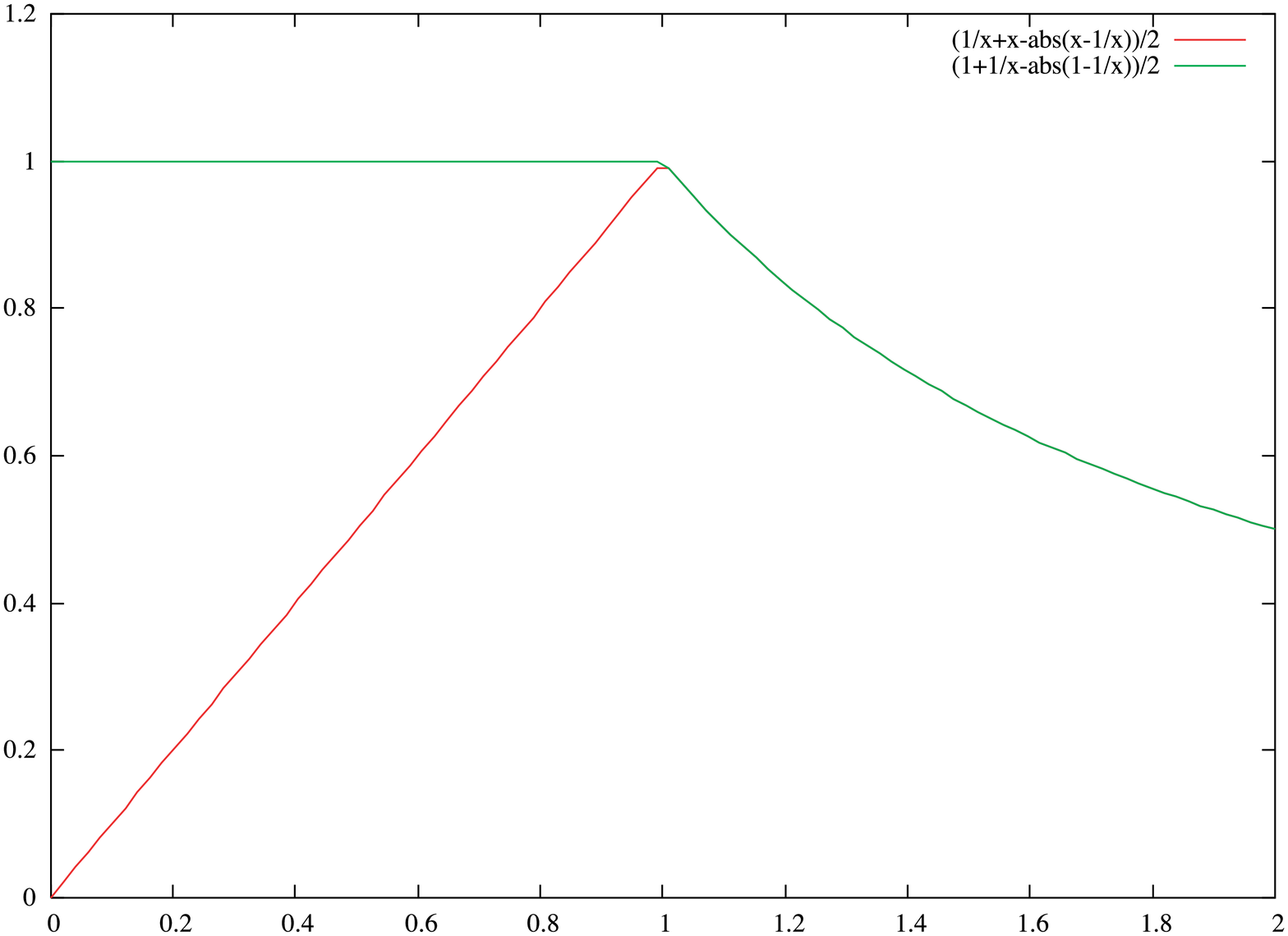}
\includegraphics[width=.49\columnwidth]{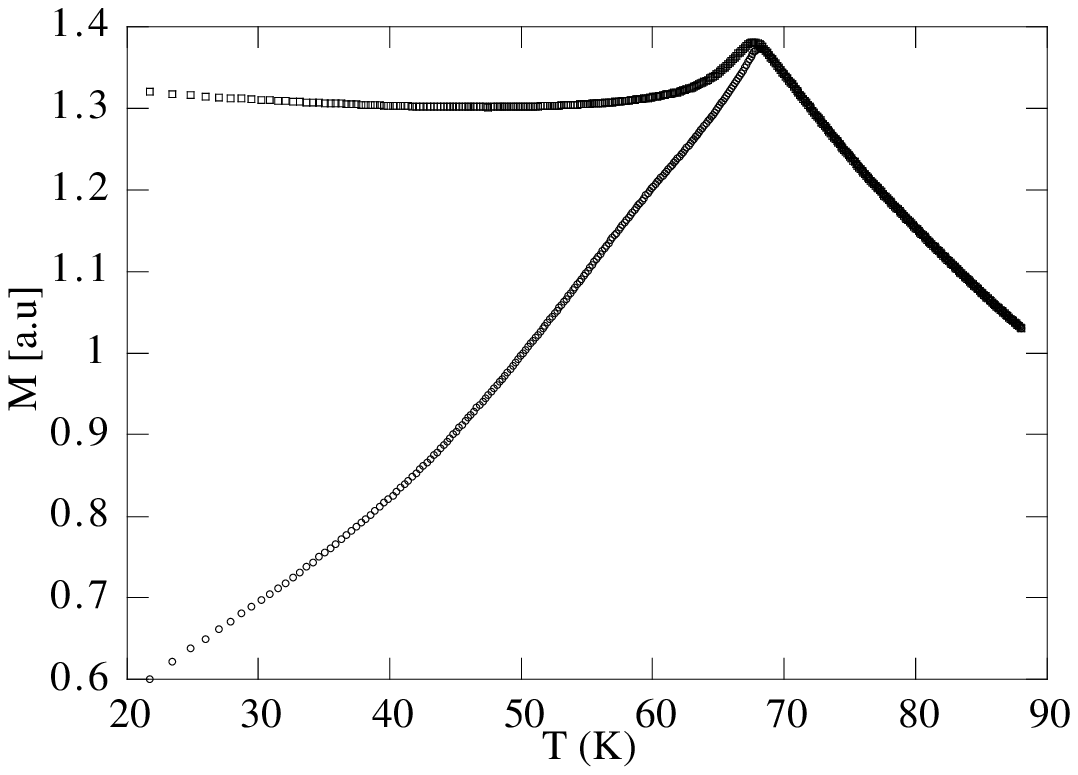}
\caption{The two susceptibilities; upper panel, the analytic results in the mean field approximation
\cite{MPV}; lower panel, the experimental results for a metallic spin glass \cite{EXP1} }
\label{TWOS}
\end{figure}

The difference of the two susceptibilities is the hallmark of replica symmetry breaking.
In fig. (\ref{TWOS}) we have both the analytic results for the SK model \cite{MPV} and  the experimental data on
metallic spin glasses \cite{EXP1}. The similarities between the two panels are striking.

Which are the differences of this phenomenon with a well known effect; i.e. hysteresis?
\begin{itemize}
	 \item Hysteresis is due to defects that are localized in space and produce a finite barrier in
	 free energy.  The mean life of the metastable states is finite and it  is roughly $\exp (\beta
	 \Delta F)$ where $\Delta F$ is a number of order 1 in natural units.

    \item In the mean field theory of spin glasses the system must cross barriers that correspond to
	 rearrangements of arbitrary large regions of the system. The largest value of the barriers diverge 
	 in the thermodynamic limit.
	 
	 \item In hysteresis, if we wait enough time the two susceptibilities  coincide, while
	 they remain always different in this new framework if the applied magnetic field is
	 small enough (non linear susceptibilities are divergent).
\end{itemize}

The difference between hysteresis and the this new picture (replica symmetry breaking) becoming clearer as we
consider fluctuation dissipation relations during aging \cite{Cuku1,Cuku2,FM,FMPP,OCIO,PNAS}, but this would be take
to far from our study of equilibrium properties.

\section{Some other applications}

Some of the physical ideas that have been developed for spin glasses have also been developed
independently by people who were working in the study of structural glasses.  However in the fields of
structural glasses there were no soluble models that displayed interesting behavior so most of the
analytic tools and of the corresponding physical insight were firstly developed for spin glasses (for a review
see for example \cite{LH2002}).

The two fields remained quite separate one from the other also because there was a widespread
believe that the presence of quenched disorder (absent in structural glasses) was a crucial
ingredient of the theory developed for spin glasses.  Only in the middle of the 90's it became clear
that this was a misconception \cite{SIN,BMEZ} and it was realized that the spin glass theory may be
applied also to systems where non-intrinsic disorder is present (e.g. hard spheres).

This became manifest after the discovery of models where there is a transition from a high temperature
\emph{liquid} phase to a low temperature glassy phase (some of these models at low temperature have two
phases, a disorder phase and an ordered, {\em crystal} phase).  Spin glass theory was able to predict the
correct behavior both at  high and at low temperature \cite{SIN,BMEZ,GRAFFI}.

There are many other problems where these ideas have been used.  In the past there has been a great interest in
applying them to problems with a biological flavor, like the computation of the optimal performance of
associative memories based on neural networks \cite{AMIT}.  Also learning abilities of simpler objects like
the perceptrons can be computed.

In recent years they have been widely used in the field of combinatorial optimization problems and 
constraint satisfaction, but this will be discussed elsewhere in this volume \cite{MM}.

\section{Conclusions}
From the retrospective point of view it may seem extraordinary the amount of new ideas that has been
necessary in order to understand in depth the basic properties of spin glasses and other glassy
system.  A strong effort should be done in order to have better quantitative predictions and a more
precise comparison between theory and experiments.  
 
I ma convinced that there are still many very interesting surprises that will come from a careful study of
spin glasses (and related problem) and it is quite likely that the most exciting result will come from an area
I am not able to indicate.  I am very curious to see the future of the field.
  
\section*{Appendix I: A useful little theorem}
In this appendix we want to prove a useful little theorem \cite{MP1Be}, which may be useful in many contests.

{\em Theorem:} consider a set of $M (\gg 1)$ 
iid random free energies $f_\alpha$, ($\alpha \in \{ 1 \ldots M \} $)
 distributed with the exponential 
density \form{ONESTEP} , and a set of 
$M$  positive numbers $a_\alpha$. Then, neglecting terms 
which go to zero when $ M$ 
goes to infinity, the following relation holds:
\bea
\lan \ln\(({\sum_\alpha a_\alpha \exp(-\beta f_\alpha) \over
\sum_\alpha  \exp(-\beta f_\alpha)} \)) \ran_{f}\\
\equiv 
\lan\ln\((\sum_\alpha  w_{\alpha} a_\alpha\)) \ran_{f}=
{1 \over x} \ln \(({1 \over M} \sum_\alpha a_\alpha^x \)) \label{THEO} \, ,
\eea
where $\lan \cdot\ran_{f}$ denotes an average over the distribution of $f$.

{\em Corollary 1:} In the same conditions as the theorem, for any set
of $M$ real numbers $b_\alpha$, one has:
\be
\left\lan{\sum_\alpha a_\alpha b_\alpha \exp(-\beta f_\alpha) \over
\sum_\alpha a_\alpha \exp(-\beta f_\alpha)} \right\ran_{f}
=
{\sum_\alpha a_\alpha^x b_\alpha  \over
\sum_\alpha a_\alpha^x }\ .
\ee

{\em Corollary 2:}
If the numbers $a_\alpha$ are $M$
 iid positive  random variables, such that the average of
$a^x$ exists, 
which are uncorrelated with the $f_\alpha$,
then one has:
\be
\left \lan \ln\(({\sum_\alpha a_\alpha \exp(-\beta f_\alpha) \over
\sum_\alpha  \exp(-\beta f_\alpha)} \)) \right\ran_{f}
\equiv 
\left\lan\ln\((\sum_\alpha  w_{\alpha} a_\alpha\)) \right\ran_{f}=
{1 \over x} \ln \(( \lan a_\alpha^x \ran_{a} \)) \label{THEO_coro} \ ,
\ee
where $\lan \cdot \ran_{a}$ denotes an average over the 
distribution of $a$.

{\em Remark:}
We notice that in the two limits $x \to 0 $ and $x\to 1$ the equations can be simply 
understood:
\begin{itemize}
    \item
    In the limit $x=0$, in a typical realization of the random  free energies,
 only one weight $w$ is equal to one and all the others are zero. 
Averaging over the 
realizations of free energies amounts to spanning uniformly the set of indices
of this special non zero weight.
\item
    In the limit $x=1$ the number of relevant $w$ goes to infinity and each 
individual 
    contribution goes to zero. An infinite number of 
    term is present in the l.h.s. of \form{THEO} and the 
    r.h.s. of the \form{THEO} becomes 
$\ln \[[ (1/M) \sum_\alpha  a_\alpha \]]$, as it should.
\end{itemize}

\end{document}